\documentclass[reviewcopy]{elsart}

\usepackage[dvips]{graphicx}
\usepackage{algorithmic}
\usepackage{algorithm}
\usepackage{bm}
\usepackage{psfrag}
\usepackage{color}
\usepackage{subfigure}

\usepackage{amsmath}
\usepackage{amssymb}

\newcommand{\bs}{\bm}
\def\RR{ \mathbb R}
\def\bz{\mathbf z}

\def\bt{\bs{\theta}}
\def\bq{\mathbf{q}}
\newcommand{\refeq}[1]{Equation (\ref{#1})}
\newcommand{\ee}{\end{equation}}
\newcommand{\be}{\begin{equation}}
\newcommand{\ec}{\end{center}}
\newcommand{\bc}{\begin{center}}
\newcommand{\eea}{\end{eqnarray}}
\newcommand{\bea}{\begin{eqnarray}}
\newcommand{\bd}{\begin{description}}
\newcommand{\ed}{\end{description}}
\newcommand{\bi}{\begin{itemize}}
\newcommand{\ei}{\end{itemize}}
\newcommand{\pa}{\partial}
\newcommand{\kl}{\operatorname{KL}}
\newcommand{\ess}[1]{\operatorname{ESS}_{#1}}

\begin{document}

\begin{frontmatter}



\title{ Free energy computations by minimization of Kullback-Leibler divergence: an 
efficient  adaptive biasing potential method for sparse representations
}

\journal{Journal of Computational Physics }
%

\author{I. Bilionis }
\ead{ ib227@cornell.edu}
\address{Center for Applied Mathematics, Cornell University, Ithaca, NY 14853, USA}

\author{P.S. Koutsourelakis \corauthref{cor2}}
\ead{pk285@cornell.edu}
\corauth[cor2]{Corresponding Author.
Tel: 607-254-5441}
\address{School of Civil and Environmental Engineering \& Center for Applied Mathematics, Cornell University, Ithaca, NY 14853, USA}

\begin{abstract}
The present paper proposes an adaptive biasing potential for the computation of free energy landscapes.
It is motivated by statistical learning arguments and unifies the tasks of biasing the molecular dynamics to escape free energy wells and estimating the free energy function, under the same objective.
It offers rigorous convergence diagnostics even though history dependent, non-Markovian dynamics are employed. It makes use of a greedy optimization scheme in order to obtain sparse representations of the free energy function which can be particularly useful in multidimensional cases. It employs embarrassingly parallelizable sampling schemes that are based on adaptive Sequential Monte Carlo and can be readily coupled with legacy molecular dynamics simulators. The sequential nature of the learning and sampling scheme enables the efficient calculation of free energy functions parametrized by the temperature. The characteristics and capabilities of the proposed method are demonstrated in three numerical examples.

\end{abstract}

\end{frontmatter}


\section{Introduction}
Free energy is a central concept in thermodynamics and in the study of several
systems in biology, chemistry and physics \cite{Chipot2006}. It represents a rigorous way to coarse-grain  systems consisting of very large numbers of atomistic degrees of freedom, to   probe states not accessible experimentally, to characterize global changes as well as investigate relative stabilities.
 In most applications, a brute-force computation based on sampling the atomistic
positions is impractical or infeasible  as the free energy barriers to overcome
are so large that    the system remains trapped in metastable free energy sets
\cite{Meirovitch2007,Rickman2002,Chipot2006,Ytreberg2006}.

Equilibrium techniques for computing free energy surfaces such as Thermodynamic
Integration \cite{Kirkwood1935a} or Adaptive Integration 
\cite{Swendsen2005,Fasnacht2004} require the simulation of very long atomistic
trajectories in order to achieve equilibrium and lack convergence diagnostics.
Techniques based on non-equilibrium path sampling
\cite{Jarzynski1997,Jarzynski1997a,Gelman1998,Hunter1993} lack adaptivity and require the user to specify a particular path on the reaction coordinate space connecting two energetically important free energy regions,
 which can be non-trivial a task \cite{E:2005}. Furthermore, sampling along these paths correctly might necessitate  advanced and quite involved techniques  \cite{Ciccotti2006a}. 
 More recently proposed adaptive biasing potential
\cite{Berg1992,Wang2001,Laio2002,Atchade,Dickson2010} and adaptive biasing
force \cite{Darve,Darve2001,Raiteri2006,Vanden-Eijnden2009,Henin2010}
techniques are capable  of dynamically utilizing information obtained from the
atomistic trajectories to bias the current dynamics in order to facilitate the
escape from 
metastable sets \cite{Lelievre2007}. They are able to  automatically discover important regions of the reaction coordinate space. Since  they rely on  history-dependent,
non-Markovian dynamics, it is not a priori clear, and in which sense, 
the system reaches a stationary state, although some work has been done along
theses lines  in  \cite{Bussi2006} for Langevin-type systems and
\cite{Marsili2006,Lelievre2007}.





We propose an adaptive biasing potential technique where  the two tasks of
biasing the dynamics and estimating the free energy landscape are unified under
the same objective of minimizing the Kullback-Leibler divergence between
appropriately selected distributions on the extended space that includes atomic
coordinates and the collective variables \cite{MARAGLIANO2006,Maragliano2008}.
This framework provides a natural way for selecting the basis functions used in
the approximation of the free energy and obtaining sparse representations which
is critical when multi-dimensional collective variables are used.
It allows the analyst to utilize and correct any prior information on the free
energy landscape and provides an efficient manner of obtaining good estimates
at various temperatures.
The scheme proposed is embarrassingly parallelizable  and relies on adaptive
Sequential Monte Carlo procedures which enable efficient sampling from the
high-dimensional and potentially multi-modal distributions of interest.


\section{Methodology - A statistical learning approach for adaptively calculating free energies}






\label{sec:adaptive}
For clarity of the presentation, we will first introduce our method for the so-called alchemical case and
generalize it  later  for the reaction coordinate case.
Consider a molecular system with generalized coordinates
$\bs{q}\in\mathcal{M}\subset \RR^N$
 following a Boltzmann-like 
distribution which in turn depends on some parameters
$\bs{z}\in\mathcal{D}\subset\RR^d$
\be
\label{eq:fe}
p(\bs{q}|\bs{z}) \propto \exp\left(-\beta V(\bs{q};\bs{z})\right)
\ee
where $V(\bs{q};\bs{z})$ is the potential energy of the system and $\beta$ plays
the role
of inverse temperature. The free
energy $A(\bs{z})$ is defined, up to an additive constant, by:
\be
A(\bs{z}) = -\beta^{-1}\int \exp\left(-\beta V(\bs{q};\bs{z})\right) d\bs{q}
\label{eq:freedef}
\ee
Our goal is to compute the function $A(\bz)$ over the
whole domain $\mathcal{D}$.

Let $\hat A(\bz;\bs{\theta})$ be an estimate of $A(\bz)$
parametrized by $\theta\in\bs{\Theta}\subset\RR^K$.
We adopt a statistical perspective of learning $A(\bz)$ from simulation data.  A popular approach to  carrying out regression tasks and functional
approximations relies on kernel models \cite{KW71}. Kernel regression models
have proven successful in high-dimensional scenaria where $d$ is in the order of
10 or 100 \cite{rel00tip,tip01spa}. The unknown function is selected
from a Reproducing Kernel Hilbert Space (RKHS) $\mathcal{H}_K$ induced by a
semi-positive definite  kernel $K(\cdot,\cdot)$. 
We adopt representations  
with respect to a kernel  function $K(\cdot , \cdot)$ \cite{KW71}:
\be
\label{eq:conv3}
\hat{A}(\bs{z}; \bs{\theta}) =\sum_{j=1}^K \theta_j K_j(\bs{z}, \bs{z}_j),
\quad \bs{z} \in \mathcal{D}
\ee
where $\bs{z}_j$ are points in $\mathcal{D}$ which are selected as described in the sequence,
In order to fix the additive constant, we select a point $\bz_0\in\mathcal{D}$
such that:
$\hat A(\bz;\bs{\theta}) = 0$. 
\footnote{This is always possible by changing the kernels
in Equation \ref{eq:conv3} to $K'_j(\bz, \bs{z}_j) =
K_j(\bz)-K_j(\bz_0, \bs{z}_j)$.} 

In relevant literature different types of kernel functions have been used such
as thin plate splines, multiquadrics, or Gaussians. While all these functions
can be employed in the framework  presented, we focus our discussion here on 
 Gaussian kernels which  also have an  intuitive parametrization with
regards to the {\em scale of variability }  of $\hat{A}$ as quantified by the
bandwidth parameters $\bs{\tau}_j=\{ \tau_{j,l}\}_{l=1}^d$ in each dimension:
\be
\label{eq:kernel}
K_j( \bs{z}) = K( \bs{z}, \bs{z}_j; \bs{\tau}_j )= \exp \{ - \sum_{l=1}^d \tau_{j,l} ( {z}_l -{z}_{j,l})^2  \}
\ee
Gaussian kernels in the context of free energy approximations have also been used in \cite{Laio2002,MARAGLIANO2006,Dickson2010}.

We define a joint probability distribution on the generalized coordinates
$\bs{q}$ and the parameters $\bz$ as follows:
\be
\label{eq:augmented}
p(\bq,\bs{z} \mid \bs{\theta}) = \frac{1}{Z(\bs{\theta})}
1_{\mathcal{D}}(\bs{z}) e^{- \beta \left(V(\bq,\bs{z})-\hat{A}(\bs{z};
\bs{\theta)} \right) }
\ee
where $1_{\mathcal{D}}(\bs{z})$ is the indicator function on $\mathcal{D}$ and 
$Z(\bs{\theta})$ is the normalization constant, i.e.:
\be
\label{eq:norm}
Z(\bt)=\int 1_{\mathcal{D}}(\bs{z}) e^{- \beta \left(V(\bq,\bs{z})-\hat{A}(\bs{z};
\bs{\theta)} \right) }d\bs{z} d\bt
\ee
It is noted that the first-order partial derivatives of the log of the normalization function give rise to the expectation of the respective kernel:
\be
\label{eq:norm1}
\frac{\pa \log Z}{\pa \theta_j}=\frac{1}{Z(\bt)} \frac{\pa Z}{\pa \theta_j}=\beta E_{p(\bs{z} \mid \bt)} \left[K_j(\bs{z}) \right] \\
\ee
whereas the second-order derivatives, produce the covariance between the kernels:
\be
\label{eq:norm2}
\begin{array}{ll}
\frac{\pa^2 \log Z}{\pa \theta_j \pa \theta_l}& =-\frac{1}{Z^2(\bt)} \frac{\pa Z}{\pa \theta_j} \frac{\pa Z}{\pa \theta_l}
+ \frac{1}{Z(\bt)} \frac{\pa^2 Z}{\pa \theta_j \pa \theta_l}  \\
& =\beta^2 E_{p(\bs{z} \mid \bt)} \left[
 (K_j(\bs{z})-E_{p(\bs{z} \mid \bt)} [ K_j(\bs{z}) ])
 (K_l(\bs{z})-E_{p(\bs{z} \mid \bt)} [ K_l(\bs{z}) ]) 
\right] \\
& = \beta^2 Cov_{p(\bs{z} \mid \bt)}\left[ K_j,K_l \right]
\end{array}
\ee
The expectations in the two equations above involve the {\em unknown} marginal density
with respect  to the parameters $\bs{z} \in \mathcal{D}$:
\be
\label{eq:augmar}
\begin{array}{ll}
p(\bs{z} \mid \bs{\theta}) & = \int p(\bq,\bs{z} \mid \bs{\theta}) ~d\bq \\
& =  \frac{1}{Z(\bs{\theta})} 1_{\mathcal{D}}(\bs{z}) e^{- \beta
\left(A(\bs{z})-\hat{A}(\bs{z}; \bs{\theta}) \right) }
\end{array}
\ee
which depends on the unknown free energy $A(\bs{z})$ of \refeq{eq:freedef}.

The key property of $p(\bs{z} \mid \bs{\theta})$ is that it reduces
to the uniform distribution for $\bz\in\mathcal{D}$ if and only if the
free energy
estimate is exact i.e.  $\hat{A}(\bs{z}; \bs{\theta}) = A(\bs{z}), ~\bs{z} \in
\mathcal{D}$.

If $\pi(\bs{z})=1_{\mathcal{D}}(\bs{z}) \frac{1}{\mid \mathcal{D} \mid }$ is the
uniform density on $\mathcal{D}$ (whose volume is denoted by $\mid \mathcal{D}
\mid$), then a natural strategy to estimate $A(\bs{z})$ is by minimizing a
distance metric between $\pi(\bz)$ and $p(\bz \mid \bs{\theta})$ in
\refeq{eq:augmar} above.
To that end, we propose employing the Kullback-Leibler ($\kl$) divergence
$\kl(\pi(\bz) \parallel p(\bz \mid \bs{\theta}))$ \cite{cov91ele}:
\be
\label{eq:kl}
\begin{array}{ll}
\kl(\pi \parallel p) & = \int \pi(\bs{z}) \log \frac{\pi(\bs{z})}{p(\bs{z} \mid
\bs{\theta})} ~d\bs{z} \\
& = \int \pi(\bs{z}) \log \pi(\bs{z}) d\bz-  \int \pi(\bs{z}) \log p(\bs{z} \mid
\bs{\theta})~d\bz \\
& = - \log \mid \mathcal{D} \mid - \int \pi(\bs{z}) \log p(\bs{z} \mid
\bs{\theta})~d\bz  \ge 0
\end{array}
\ee
The latter is not a metric in the mathematical sense, but it is frequently
used as a measure of the distance between two probability distributions.
It is
 always non-negative and becomes zero if and only if $\pi(\bz) \equiv p(\bs{z}
\mid \bs{\theta})$ or equivalently $\hat{A}(\bs{z}; \bs{\theta}) = A(\bs{z}),
~\bs{z} \in \mathcal{D}$ \footnote{Of interest are free-energy {\em differences}
and therefore perturbations of  $A(\bz)$ or $\hat{A}(\bs{z}; \bs{\theta})$  by a
constant are ignored}.
The aforementioned formulation offers a clear strategy for estimating  the
free energy by minimizing the following form with respect to $\bs{\theta}$:
\be
\label{eq:iota}
I(\bs{\theta})= - \int   \pi(\bs{z}) \log p(\bs{z} \mid \bs{\theta}) d\bs{z} 
\ee
Since the $\kl$-divergence is always non-negative, the formulation above provides a
lower bound on the objective function $I(\bt)$:
\be
\label{eq:iota1}
I(\bs{\theta}) \ge \log \mid \mathcal{D} \mid
\ee
which can be readily calculated and be used to monitor convergence as well as the quality of the approximation obtained.

Even though $I(\bt)$ depends on the unknown free energy $A(\bs{z})$ (from \refeq{eq:augmar}):
\be
\label{eq:iota2}
\begin{array}{ll}
 I(\bt)& =- \int   \pi(\bs{z}) \log p(\bs{z} \mid \bs{\theta}) d\bs{z}  \\
&=\beta \int \pi(\bs{z}) \left(A(\bs{z})-\hat{A}(\bs{z} ; \bt) \right)d\bs{z}+\log Z(\bt)
\end{array}
\ee
its partial derivatives $\bs{J}(\bt)=\frac{ \pa I(\bs{\theta}) }{\pa \bt}$ do not
, i.e. from \refeq{eq:norm1}:
\be
\label{eq:mini}
\begin{array}{ll}
{J}_j(\bt) & = \frac{\pa I(\bt) }{\pa \theta_j } \\
& = -\beta E_{\pi(\bz)}\left[ \frac{\pa \hat{A}}{\pa \theta_j} \right] +\frac{\pa \log Z}{\pa \theta_j} \\
&  = -\beta \left(
E_{\pi(\bz)}\left[ K_j(\bs{z}) \right] - E_{p(\bz \mid
\bs{\theta})}\left[ K_j(\bs{z})  \right] \right)
\end{array}
\ee
where $E_{\pi(\bz)}[\cdot]$  implies an  expectation with regards to $\pi(\bz)$. 

It is important to note that according to \refeq{eq:norm2}, 
the Hessian of the objective function $\bs{H}(\bt)=\frac{\pa^2 I(\bt)}{\pa \bt  \pa \bt^T}$ is proportional to the covariance between 
the kernels i.e.:
\be
\label{eq:hessian}
\begin{array}{ll}
\frac{ \pa^2 I}{\pa \theta_j \pa \theta_l} & =\frac{\pa \log Z(\bt)}{\pa \theta_j \pa \theta_l}  \\
& = \beta^2 Cov_{p(\bs{z} \mid \bt)}[K_j,K_l]
\end{array}
\ee
Hence the objective function is convex with respect to $\bt$ and there is a unique minimum.

Furthermore, the approximation of the free  energy
$\hat{A}(\bz ; \bt)$,  biases the potential of $p(\bq,\bz \mid \bt)$
(\refeq{eq:augmented}) and allows the system to overcome free energy barriers
\cite{Marsili2006}.
As in \cite{Dickson2010}, no binning is needed and the bias potential is nonlocal, providing
information about the free energy landscape not only at the states visited but
in their neighborhood as well.  
In contrast to other adaptive schemes, the proposed formulation connects the 
problems  of estimating the free energy landscape and steering the atomistic 
dynamics beyond metastable wells, under a unified umbrella, and provides a clear  convergence criterion \cite{Lelievre2007}.

From an algorithmic point of view, the proposed strategy poses two problems. The first involves 
 the selection of the kernels  to be used in the expansion of \refeq{eq:conv3}.  This is critical  to the  sparseness of the representation obtained, particularly for multidimensional $\bs{z}$.  To that end, a greedy selection   strategy is discussed in section \ref{sec:kernel} which progressively adds kernels (i.e. increases the cardinality $K$ of the expansion in \refeq{eq:conv3}) as needed.
The second problem involves the optimization of the objective function $I(\bt)$ which depends on the unknown free energy $A(\bs{z})$ and the intractable partition function $Z(\bt)$ (\refeq{eq:iota2}).
An obvious approach is by gradient descent which is discussed  in detail in section \ref{sec:opti}. This  requires the computation of expectations with respect to $p(\bs{z} \mid \bt)$ (\refeq{eq:mini}). The intractability of $p(\bs{z} \mid \bt)$ necessitates the use of Monte Carlo sampling which must be carried out in the expanded space with respect to the joint density $p(\bs{q,z} \mid \bt)$ (\refeq{eq:augmented}). This  should nevertheless be able to capture multiple modes in the high-dimensional state space consisting of atomic degrees of freedom $\bs{q}$ and parameters $\bs{z}$. 
To that end we  propose performing this step by using  non-equilibrium
path sampling techniques based on {\em adaptive Sequential Monte Carlo} schemes
discussed in section \ref{sec:smc}. Similar schemes for creating system replicas in parallel have been 
employed in \cite{Raiteri2006,Lelievre2007}. We discuss a novel adaptive version that retains previous advantages  while providing accurate  estimates at reduced computational effort. These estimates  can be readily updated as $\bt$ changes after each optimization step.

A discussion of each of the aforementioned algorithmic modules is 
contained in the ensuing sub-sections. The steps of the scheme proposed
can be found in Algorithm \ref{alg:main}.

\subsection{Optimization with noisy gradients}
\label{sec:opti}
We propose employing a gradient descent scheme  in order to determine  $\bt$, although more involved procedures such as Improved Iterative Scaling  \cite{ber97imp,pie97ind}, noisy conjugate gradients \cite{sch02tow} can be also be employed. Second-order (quasi-)Newton techniques are also possible although the unavoidable Monte Carlo noise in the computation of the Hessian (i.e. the covariance in \refeq{eq:hessian}) can destroy its positive definiteness.

Let $\bt_K$ denote the vector of kernel amplitudes
(\refeq{eq:conv3}) when $K$ such kernels are used. Let also $\bt_{K,m}$ denote
the estimate of $\bt_K$ after $m$ iterations of the gradient descent algorithm.
Then at the $(m+1)-$iteration, the following update equation could be used:
\be
\label{eq:gd}
\bt_{K,m+1}=\bt_{K,m}-\lambda \bs{J}(\bt_{K,m})\ee
where $\lambda $ is the learning rate.

Since only a noisy Monte Carlo estimate  of the gradients 
$\bs{J}(\bt)$ (\refeq{eq:mini})
is available, it is anticipated that the noise  could impede convergence.
For that purpose we propose employing a stochastic approximation variant of the Robbins \& Monro scheme \cite{rob51sto,cap05inf}.   Rather than  increasing the simulation
size in order to reduce the variance, we compute a weighted average of the gradient's estimates at the current and previous iterations.   By employing  a decreasing sequence of weights, 
information from the earlier iterations gets discarded gradually and more emphasis is placed on the 
recent iterations. As it is shown in \cite{del99con}, this method converges with a fixed sample size. In particular, if $\hat{\bs{J}}(\bt_{K,m})$ denotes the Monte Carlo estimate of the gradient (the details of this estimator are discussed in section \ref{sec:smc}) at the $m^{th}$ iteration, then we calculate:
\be
\label{eq:saem}
\tilde{\bs{J}}_{m}=(1-\eta_{m})\tilde{\bs{J}}_{m-1}+\eta_{m}\hat{\bs{J}}(\bt_{K,m})
\ee
and, rather than \refeq{eq:gd}, we update  $\bt$ as follows:
\be
\label{eq:rm}
\bt_{K,m+1}=\bt_{K,m}-\lambda \tilde{\bs{J}}_{m}
\ee
where the sequence of weights $\{\eta_m\}$ is such that   $\sum_{m=1}^\infty\eta_m = \infty$ 
and $\sum_{m=1}^\infty\eta_m^2 < \infty$\footnote{
A family of such sequences that was used in this work is $\eta_m = \eta m^{-p}$ with $1/2 < p \le 1$.
}.

\subsection{Kernel selection - Sparse representation of free energy
landscape}
\label{sec:kernel}

 A critical objective  in the proposed framework relates to the {\em sparseness}
of the free energy approximation i.e. the cardinality $K$ of the expansion in
\refeq{eq:conv3}. This is important in at least two ways. Firstly, because
sparser representations can more clearly expose salient features of the free
energy landscape, and as a consequence, of the atomistic ensemble considered.
Secondly, because they reduce the number of parameters $\bt$ with respect to
which the optimization problem needs to be solved (section \ref{sec:opti}). 
Given a vocabulary of potentially overcomplete basis functions and a prescribed
$K$, the problem amounts to identifying those kernels (\refeq{eq:conv3}) that
best approximate the true free energy surface i.e. minimize the $\kl$ divergence
for $\bs{z} \in \mathcal{D}$ (\refeq{eq:kl}). This obviously implies an
excessive computational effort since the aforementioned optimization problem
would need to be solved  for all possible $K-$sized combinations of basis
functions. 

For that purpose, we propose a  hierarchical scheme that proceeds by adding
a single kernel at each step. Similar greedy optimization procedures have been
successfully applied in maximum entropy  problems \cite{pie97ind}. 
Without loss of generality, one can consider a vocabulary of functions that
consists of the isotropic Gaussian kernels discussed in \refeq{eq:kernel}. Each
of these is parametrized by the location $\bz_j$ of the kernel and its
bandwidth $\bs{\tau}_j$. 
Given $K$ such kernels, the corresponding parameters $\bt_K=\{ \theta_j
\}_{j=1}^K$ (\refeq{eq:conv3}) that 
minimize $I(\bt)$ in \refeq{eq:iota} and samples from the density $p(\bz \mid
\bt_K)$ (\refeq{eq:augmar} or \refeq{eq:augmented}), we propose selecting the
$(K+1)^{th}$ kernel by choosing $\left( \bz_{K+1}, \bs{\tau}_{K+1}=\{\tau_{K+1,l}\}_{l=1}^d  \right)$ that maximize:
\be
\label{eq:add_kernel}
(\bz_{K+1}, \bs{\tau}_{K+1})= arg \max_{(\bz_{K+1}, \tau_{K+1})}  \left|
E_{\pi(\bz)}\left[ K(\bz, \bz_{K+1}; \tau_{K+1})  \right] - E_{p(\bz \mid
\bt_{K})}\left[  K(\bz, \bz_{K+1}; \tau_{K+1})\right] \right|
\ee
Based on \refeq{eq:mini}, this suggests augmenting our expansion with the
kernel that locally maximizes the 
gradient of  $I(\bt)$.   Intuitively this means that we incorporate the kernel
function whose expected value with respect to the target, uniform distribution
is worst approximated by the current density  $p(\bz \mid \bt_{K})$.
 It is obviously a suboptimal strategy, that is necessitated by  reasons of
computational cost. The maximization of the objective in \refeq{eq:add_kernel}
can be readily carried out given samples from $p(\bz \mid \bt_{K})$. The same
formulation  can be applied to any type of kernel or overcomplete basis employed
(e.g. wavelets). 
The proposed strategy promotes  sparseness and computational efficiency while
offering a progressive resolution of the free energy landscape that naturally
involves kernels with  larger bandwidths (smaller $\bs{\tau}$) in the first steps,
and successive unveiling of the finer details which can be captured by kernels
of smaller bandwidths (i.e larger $\bs{\tau}$).

Furthermore,  it offers a rigorous metric for monitoring convergence. In
particular if $\hat{A}_{K}(\bs{z}; \bt_K)$ and  $\hat{A}_{K+1}(\bs{z};
\bt_{K+1})$ denote the free-energy approximations obtained at successive steps
using $K$ and $K+1$ kernels respectively, and $p_K$ and $p_{K+1}$ the
corresponding densities in \refeq{eq:augmented}, then the improvement in terms
of Kullback-Leibler divergence (\refeq{eq:kl}), denoted by $\Delta_{K+1}$ can be
assessed with:
\be
\label{eq:kl_impr}
\begin{array}{ll}
 0 \le \Delta_{K+1} & = \kl(\pi \parallel p_K)- \kl(\pi \parallel p_{K+1}) \\
& = -\beta E_{\pi}\left[ \hat{A}_{K+1}(\bs{z}; \bt_{K+1})-\hat{A}_{K}(\bs{z};
\bt_K) \right]+ \log \frac{ Z(\bt_{K+1})}{ Z(\bt_K) }
\end{array}
\ee
The expectation with respect to the uniform can in general be calculated
analytically whereas the ratio of normalizing constants $\log \frac{
Z(\bt_{K+1})}{ Z(\bt_K) }$ (\refeq{eq:augmented}) is a direct output of the
Sequential Monte Carlo sampling that is used to sample from the augmented
densities and  is discussed in the next section.

\subsection{Adaptive Sequential Monte Carlo}
\label{sec:smc}
The learning scheme proposed relies on efficient computations of the gradients
appearing in \refeq{eq:mini}.  These depend on   expectations with respect to
$p(\bz \mid \bs{\theta})$ (\refeq{eq:augmar}) which is not available
analytically  since the actual free energy $A(\bz)$ is unknown. We resort to a
Monte Carlo scheme that   draws samples from the joint density $p(\bq, \bz \mid
\bt)$ in \refeq{eq:augmented} which involves the atomic degrees of freedom
$\bq$. 
 A brute-force approach would generally be inefficient as simulating atomistic
trajectories suffers from well-known difficulties such as the
high-dimensionality of $\bq$, the disparity in scales between $\bq$ and $\bz$
and the presence of several metastable wells \cite{Smit2002}. Furthermore, since
$p(\bq, \bz \mid \bt)$ depends on $\bt$, new samples would have to drawn every time
$\bt$ changes after each  iteration of the optimization routine.

For these reasons, we propose a parallelizable strategy that relies on
Sequential Monte Carlo samplers (SMC,  \cite{mor06seq,mor04fey}). These offer a
general statistical  perspective that unifies a range of pertinent schemes that
have been proposed in the context of non-equilibrium path sampling following
Jarzynski's work \cite{Jarzynski1997,Stoltz}.
We propose novel extensions that allow the algorithm to automatically adapt to
the difficulties of the target density. They retain the  ability to interact
seamlessly with legacy, molecular dynamics simulators.

 The proposed SMC schemes offer a flexible framework  for sampling from a {\em
sequence of unormalized  probability distributions}  and are therefore highly suited for the 
dynamic setting  of the problem at hand where the target density $p(\bq, \bz
\mid \bt)$ changes with $\bt$. For a given $\bt$, they approximate $p(\bq, \bz
\mid \bt)$  with a  set of $n$ random samples (or {\em particles})  $\{
\bq^{(i)}, \bz^{(i)} \}_{i=1}^n$, which  are  updated using a combination of
{\em importance sampling},  {\em resampling}  and MCMC-based {\em rejuvenation}
mechanisms \cite{del06seq}. Each of these particles is associated with an {\em
importance  weight} $w^{(i)}$ 
 which is proportional to $p(\bq^{(i)}, \bz^{(i)} \mid \bt)$. The weights are updated sequentially along with the particle locations in order to provide a particulate
approximation:
\be                                                                            
\label{eq:smc1}
p(\bq, \bz \mid \bt)  \approx \sum_{i=1}^n ~W^{(i)}~ \delta_{\bq^{(i)}} (\bq) 
\delta_{\bz^{(i)}} (\bz)
\ee
where $W^{(i)}=w^{(i)}/\sum_{k=1}^N w^{(k)}$ are the normalized weights and  $\delta_{\bs{x}^{(i)}}(.)$ is the Dirac function centered at
$\bs{x}^{(i)}$. These particles and weights can be used to estimate expectations
of any $p(\bq, \bz \mid \bt)$-integrable function \cite{mor04fey,Chopin:2004}. In particular for  \refeq{eq:mini}: 
\be
\label{eq:smc2}
 \sum_{i=1}^n ~W^{(i)}~K_j(\bz)  \rightarrow \int K_j(\bz)~p(\bq,
\bz \mid \bt)~d\bq d\bz=E_{p(\bz \mid \bs{\theta})}\left[ K_j (\bz)
\right] \quad \textrm{\em (almost surely)}
\ee

The proposed SMC algorithms will be used iteratively, after each step of the
gradient descent algorithm. Given two successive estimates  $\bt_{K,m}$ and
$\bt_{K,m+1}$ (\refeq{eq:rm}) and a particulate approximation of $p(\bq, \bz
\mid \bt_{K,m})$, the goal is to obtain new samples from $p(\bq, \bz \mid
\bt_{K,m+1} )$ (Algorithm \ref{alg:main}) and compute the new expectations in
\refeq{eq:mini} based on \refeq{eq:smc2}.  The
quality of the Monte Carlo estimates in \refeq{eq:smc2}  depends on the
proximity of the distributions $p(\bq, \bz \mid \bt_{K,m})$ and $p(\bq, \bz \mid
\bt_{K,m+1})$. We propose building a path of intermediate,
 unormalized distributions that will bridge this gap  based on
 \refeq{eq:augmented} \footnote{subscripts $K$ and $m$ indicating the number of kernels and optimization iterations respectively have been dropped}:
\be
\label{eq:interm}
\begin{array}{ll}
\pi_{\gamma}(\bq,\bz) & \propto  p(\bq,
\bz \mid (1-\gamma)  \bt_{K,m}+\gamma  \bt_{K,m+1}) \\
& =\exp \left\{ -\beta \left( V(\bq,\bz)-\hat{A}(\bz; \bt_{\gamma}) \right) \right\}, \quad  \gamma \in [0,1] 
\end{array}
\ee
where:
\be
\label{eq:interm1}
\bt_{\gamma}=(1-\gamma)  \bt_{K,m}+\gamma  \bt_{K,m+1}
\ee
Clearly for $\gamma=0$ one recovers $p(\bq,\bz \mid \bt_{K,m})$ and for $\gamma=1$, $p(\bq,\bz \mid \bt_{K,m+1}$). The role of these auxiliary distributions is to provide a smooth transition path where importance sampling can be efficiently applied. Naturally, the more intermediate distributions are
considered along this path, the higher the accuracy of the final estimates, but
also the higher the  computational cost. 
On the other hand too few intermediate distributions $\pi_{\gamma}$ can adversely affect the overall accuracy of the approximation.

To that end we propose an {\em adaptive} SMC scheme that  automatically determines
 the number of intermediate distributions needed \cite{del06seq,kou09acc}.  In
this process we are guided by the   Effective Sample Size (ESS, \cite{liu01mon}). 
In particular, let $S$  be the total number of intermediate distributions 
(which is unknown a priori) and $\gamma_s, ~s=1,2,\ldots, S$  
the associated reciprocal temperatures such that
 $0=\gamma_1< \gamma_2 < \ldots < \gamma_S=1$, which are 
 also unknown a priori. Let also $\{(\bq^{(i)}_s,\bs{z}_s^{(i)}),~W_s^{(i)} \}_{i=1}^N$ 
 denote the particulate approximation of $\pi_{\gamma_s}$ defined as in 
 \refeq{eq:interm} for $\gamma=\gamma_s$. 
 The Effective Sample Size of these particles is then defined as   
 $\ess{s}=1/\sum_{i=1}^N (W_s^{(i)})^2$ and provides a measure of 
 the population variance. One extreme, i.e. when $\ess{s}=1$, 
 arises when a single particle has a  unit  normalized weight whereas 
 the rest have zero weights and  as a result provide no information.
 The other extreme, i.e. $\ess{s}=N$, arises when all the particles 
 are  equally  informative and have equal weights $W_s^{(i)}=1/N$.  
 
If  the next bridging distribution $\pi_{\gamma_{s+1}}$  is very
 similar to $\pi_{\gamma_{s}}$ (ie. $\gamma_{s+1} \approx \gamma_s$), then  $\ess{s+1}$ 
 should not be that much different from $\ess{s}$. On the other hand if that difference 
 is pronounced then $ESS_{s+1}$ could drop dramatically. 
 Hence in determining the next auxiliary distribution, we define an 
 acceptable reduction in the $\ess{}$, i.e. $\ess{s+1} \ge \zeta ~\ess{s}$ 
 (where $\zeta<1$) and prescribe $\gamma_{s+1}$ (\refeq{eq:interm})  accordingly.

 \begin{algorithm}[!h]                      
\caption{Adaptive SMC}          
\label{alg:asmc}                           
\begin{algorithmic}                    
\REQUIRE{$s=1$ and $\gamma_1=0$ and a population $\{(\bq_{1}^{(i)},\bs{z}_1^{(i)}),~w_{1}^{(i)} \}_{i=1}^N$ which approximate $\pi_{\gamma_1} \equiv p(\bq,\bz \mid \bt_{K,m})$ in \refeq{eq:interm}}
\ENSURE{
The final population $\{(\bs{\theta}_{s}^{(i)}, \bs{d}_{s}^{(i)}), w_{s}^{(i)} \}_{i=1}^N$ provides a particulate approximation of $\pi_{\gamma_s}$  in the sense of Equations        (\ref{eq:smc1}), (\ref{eq:smc2}).
}
\WHILE{$\gamma_s<1$}
\STATE{$s \gets s+1$}
\STATE\COMMENT{{\bf Reweighting-Importance Sampling}}
\STATE{
Let
\be
\label{eq:weights}
\begin{array}{ll}
w_{s}^{(i)}(\gamma_s) & =w_{s-1}^{(i)}~\frac{ \pi_{\gamma_{s}}(\bq_{s-1}^{(i)},\bs{z}_{s-1}^{(i)}) }{ \pi_{\gamma_{s-1}}( \bs{q}_{s-1}^{(i)}, \bs{z}_{s-1}^{(i)}) } \\
& = w_{s-1}^{(i)}~ \frac{ \exp \left\{-\beta(V(\bs{q}_{s-1}^{(i)}, \bs{z}_{s-1}^{(i)})-\hat{A}(\bs{z}_{s-1}^{(i)}; \bt_{\gamma_s}) \right\} }{
\exp \left\{-\beta(V(\bs{q}_{s-1}^{(i)}, \bs{z}_{s-1}^{(i)})-\hat{A}(\bs{z}_{s-1}^{(i)}; \bt_{\gamma_{s-1}}) \right\} } \\
& =w_{s-1}^{(i)}~\exp \left\{-\beta( \hat{A}(\bs{z}_{s-1}^{(i)}); (\gamma_{s}-\gamma_{s-1})(\bt_{K,m+1}-\bt_{K,m}) \right\}
\end{array}
\ee
be the \emph{updated} weights as a function of $\gamma_s$.
Determine $\gamma_s\in(\gamma_{s-1},1]$ so that 
$\ess{s} = \zeta \ess{s-1}$ 
. 
}
\STATE\COMMENT{{\bf Resampling}}
\IF{$\ess{s} \le \ess{min}$}
\STATE Resample
\ENDIF
\STATE\COMMENT{{\bf Rejuvenation}}
\STATE{
Use an MCMC kernel $P_{s}\left( (\bs{q}_{s-1}^{(i)}, \bs{z}_{s-1}^{(i)}), (\bs{q}_{s}^{(i)}, \bs{z}_{s}^{(i)})\right)$ that leaves $\pi_{\gamma_s}$
invariant to perturb each particle $(\bs{q}_{s-1}^{(i)}, \bs{z}_{s-1}^{(i)}) \to (\bs{q}_{s}^{(i)}, \bs{z}_{s}^{(i)})$.
}
\ENDWHILE
\end{algorithmic}
\end{algorithm}

The proposed adaptive SMC algorithm is summarized in Algorithm \ref{alg:asmc}.
It should be noted that unlike MCMC schemes, the particle perturbations in the {\em Rejuvenation} step do not require that the $P_{s}(.,.)$ is {\em ergodic} \cite{mor06seq}. It suffices that it is a $\pi_{\gamma_s}$-invariant kernel, which readily allows adaptively changing its parameters in order to achieve better mixing rates. In the examples presented a Metropolized Gibbs scheme  was used to sample $\bs{q}$ and $\bs{z}$ separately by employing a Metropolis-Adjusted Langevin Algorithm (MALA) for each set of coordinates \cite{rob04mon}.
In particular given $\left( \bs{q}_{s-1}^{(i)}, \bs{z}_{s-1}^{(i)} \right)$ these consist of:
\bi
\item Updating $\bs{q}_{s-1}^{(i)} \to \bs{q}_{s}^{(i)}$:
\be
\label{eq:malaq}
\begin{array}{ll}
\bs{q}_{s}^{(i)}-\bs{q}_{s-1}^{(i)} & =\frac{\Delta t_q}{2} \nabla_{\bq} \pi_{\gamma_s}(\bs{q}_{s-1}^{(i)}, \bs{z}_{s-1}^{(i)} ) +\sqrt{\Delta t_q} \bs{r}_q \\
& = -\frac{\beta \Delta t_q}{2} \nabla_{\bq} V(\bs{q}_{s-1}^{(i)}, \bs{z}_{s-1}^{(i)}) +\sqrt{\Delta t_q} \bs{r}_q 
\end{array}
\ee
\item Updating $\bs{z}_{s-1}^{(i)} \to \bs{z}_{s}^{(i)}$:
\be
\label{eq:malaz}
\begin{array}{ll}
\bs{z}_{s}^{(i)}-\bs{z}_{s-1}^{(i)} & =\frac{\Delta t_z}{2} \nabla_{\bq} \pi_{\gamma_s}(\bs{q}_{s}^{(i)}, \bs{z}_{s-1}^{(i)} ) +\sqrt{\Delta t_z} \bs{r}_z \\
& = -\frac{\beta \Delta t_z}{2} \left( \nabla_{\bz} V(\bs{q}_{s-1}^{(i)}, \bs{z}_{s-1}^{(i)}) -\nabla_{\bz}\hat{A}(\bs{z}_{s-1}^{(i)}; \bt_{\gamma_s}) \right)+\sqrt{\Delta t_q} \bs{r}_q 
\end{array}
\ee
\ei
where $\bs{r}_q$ and $\bs{r}_z$ are i.i.d standard Gaussian vectors. 
A Metropolis accept/reject step with respect to the target invariant density $\pi_{\gamma_s}(.)$ was performed after each update. 
Two different time steps were used $\Delta t_q$ and $\Delta t_z$ for the $\bq$ and $\bz$ coordinates respectively. Their values were adjusted after each iteration $s$ so as to retain an average acceptance ratio (over all particles $n$) between $50\%$ and $80 \%$ \cite{rob01opt}.
The adaptivity afforded by the proposed scheme relies on the fact that ergodicity is not required from the rejuvenation step. As a result several MALA time steps can be performed in Equations (\ref{eq:malaq}) and (\ref{eq:malaz}) (at additional computational expense) or other molecular dynamics samplers can be employed which could potentially exhibit better mixing or fit more closely to the physics of the problem at hand \cite{Cances2005}.

Finally we note that the estimates of the ratio of normalization constants $Z_s /Z_{s-1}$ between two successive unormalized densities $\pi_{\gamma_{s-1}}$ and $\pi_{\gamma_s} $can be obtained by averaging the unormalized updated weights in \refeq{eq:weights} as a direct consequence of the importance sampling identity:
\be
\label{eq:norm_ratio}
\begin{array}{ll}
\frac{Z_s}{Z_{s-1}} &=\frac{ \int \pi_{\gamma_s}(\bq,\bz) ~d\bq d\bz }{  \int \pi_{\gamma_{s-1}}(\bq,\bz) ~d\bq d\bz } \\
& = \int \frac{\pi_{\gamma_s}(\bq,\bz) }{ \pi_{\gamma_{s-1}}(\bq,\bz) } \frac{\pi_{\gamma_{s-1}}(\bq,\bz)}{Z_{s-1}}  ~d\bq d\bz  \\
& \approx \sum_{i=1}^n W_{s-1}^{(i)}   \frac{\pi_{\gamma_s}(\bq_{s-1}^{(i)},\bz_{s-1}^{(i)}) }{ \pi_{\gamma_{s-1}}(\bq_{s-1}^{(i)},\bz_{s-1}^{(i)}) } 
\end{array}
\ee
These estimators can be telescopically multiplied (\cite{mor06seq,kou08des}) in order to compute the ratio of normalization constants between any pair of distributions as required in \refeq{eq:kl_impr}.

Given the preceding discussion in sections \ref{sec:opti}, \ref{sec:kernel} and \ref{sec:smc}, we summarize below the basic steps in the proposed free energy computation scheme:
In the inner loop and for fixed $K$, gradient descent (subsection \ref{sec:opti}) is performed which makes use of the adaptive SMC scheme (subsection \ref{sec:smc}) in order to compute the expectations in the gradient. In the outer loop, the cardinality of the expansion $K$ is increased by adding one  kernel  (i.e. $ K \leftarrow  K+1$) based on \refeq{eq:add_kernel}. This is terminated when the $\kl$ gain (\refeq{eq:kl_impr}) does not exceed a prescribed tolerance. 

\begin{algorithm}
\caption{Calculation of the free energy at a given temperature.}
\label{alg:main}
\begin{algorithmic}
\REQUIRE{
$K=0$, $\bt_0 \equiv 0$ and a particulate
approximation of $p(\bq, \bz \mid \bt_0 )$ (\refeq{eq:augmented}) at the desired temperature $\beta$.
}
\WHILE \TRUE
	\STATE{Calculate $\Delta_K$ based on \refeq{eq:kl_impr}.}
	\IF{$\Delta_K \le \text{tol}$}
		\STATE Break the loop.
	\ELSE
	\STATE{Add the  optimal $(K+1)^{th}$ kernel based on \refeq{eq:add_kernel} and set $K \leftarrow K+1$}
	\REPEAT
		\STATE{Estimate gradient at $\bt_{K,m}$ and calculate update $\bt_{K,m+1}$  based on \refeq{eq:rm}}
		\STATE{Use adaptive SMC (section \ref{sec:smc}) to construct particulate approximation of $p(\bq, \bz \mid \bt_{K,m+1})$ from  $p(\bq, \bz \mid \bt_{K,m})$.}
	\UNTIL{Convergence criteria are met.}
	\ENDIF
\ENDWHILE

\end{algorithmic}
\end{algorithm}

\subsection{Obtaining the free energy landscape for various temperatures.}
\label{sec:beta}
The methodology described in the previous sections is suitable for calculating 
the free energy as a function of $\bz$ at a given temperature. However, one
is often interested in the free energy 
landscape as a function of the temperature also. 
In order to achieve this goal  we make use of the following two facts.  Firstly,
the free energy landscape at higher temperatures is 
flatter  and secondly that nearby temperatures have similar free energies landscapes.
Based on these, we propose a natural extension to the sequential sampling framework of subsection \ref{sec:smc} that can efficiently  produce estimates of the  free energy 
 at various temperatures.
The idea is to start from a higher temperature, compute
the free energy as described before, then gradually move
towards lower temperatures using the free energy
of the previous temperature as an initial guess for the new one.
In particular given the free energy estimate  $\hat{A}_{\beta_1}(\bs{z} ; \bt(\beta_1))$  and the particulate approximation of the joint density $p_{\beta_1}(\bq,\bz \mid \bt(\beta_1))$ at a temperature $1/\beta_1$, we propose employing the aforementioned adaptive SMC in order to obtain a particulate approximation of the following joint density at $\beta_2 > \beta_1$ (i.e. for lower temperature):
\be
\label{eq:densb}
p_{\beta_2}(\bq,\bz \mid \bt(\beta_1)) \propto \exp \left\{- \beta_2 \left( V(\bq,\bz) -\hat{A}_{\beta_1}(\bs{z} ; \bt(\beta_1))  \right) \right\}
\ee
The iterations enumerated in Algorithm \ref{alg:main} can then be carried out in the same fashion by updating the existing $\bt$ as well as adding new kernels if the convergence criteria are not satisfied.

The critical step involves building a sequence of distributions from $p_{\beta_1}(\bq,\bz \mid \bt(\beta_1))$ to $p_{\beta_2}(\bq,\bz \mid \bt(\beta_1))$ in \refeq{eq:densb}. For this purpose 
 and similarly to a simulated annealing schedule we employ:
\be
\label{eq:interm2}
\pi_{\gamma}(\bq,\bz) \propto \exp \left\{- ((1-\gamma)\beta_1+\gamma \beta_2) \left( V(\bq,\bz) -\hat{A}_{\beta_1}(\bs{z} ; \bt(\beta_1))  \right) \right\}
\ee
The steps in Algorithm \ref{alg:asmc} should be adjusted to the aforementioned sequence of bridging distributions with the most striking difference in the {\em Reweighing} step where the updated weights in \refeq{eq:weights} should now be given by:
\be
\label{eq:weights2}
\begin{array}{ll}
w_{s}^{(i)}(\gamma_s) & =w_{s-1}^{(i)}~\frac{ \pi_{\gamma_{s}}(\bq_{s-1}^{(i)},\bs{z}_{s-1}^{(i)}) }{ \pi_{\gamma_{s-1}}( \bs{q}_{s-1}^{(i)}, \bs{z}_{s-1}^{(i)}) } \\
& =w_{s-1}^{(i)}~\exp \left\{-(\gamma_{s}-\gamma_{s-1})(\beta_2-\beta_1) \left( V(\bq_{s-1}^{(i)},\bs{z}_{s-1}^{(i)}) - \hat{A}_{\beta_1}(\bs{z} ; \bt(\beta_1)) \right) \right\}
\end{array}
\ee
We demonstrate the efficacy of such an approach in the last example of section \ref{sec:examples}. It is finally noted that at the beginning of iterations at each new temperature, kernels with very small weights $\theta_j$ were removed if $\frac{\theta_j }{ \max_i \theta_i} \le 0.01$.

\subsection{The reaction coordinate case.}
\label{sec:rc}
The proposed method was described for the alchemical case. However, it
is straightforwardly generalized to cover also the general reaction
coordinate case. Let $\bs{\xi}:\mathcal{M}\rightarrow\mathcal{D}$ be a function
of the system coordinates. 
This function is called a reaction coordinate \cite{lel10fre}.
It is evident that $\bs{q}$ can be viewed in a probabilistic framework
as a random variable and so: 
\be
\label{eq:rcrv}
\bz = \bs{\xi}(\bs{q})
\ee
is also a random variable. The probability distribution of $\bz$ can be
found by integrating out the $\bs{q}$:
\be
\label{eq:rcpdf}
p(\bz \mid \beta) = \int p(\bs{q}) \delta(\bs{\xi}(\bs{q})-\bz)d\bs{q} 
\propto 
\int \exp\left(-\beta V(\bs{q})\right)\delta(\bs{\xi}(\bs{q})-\bz)d\bs{q}
\ee
The free energy $A(\bz)$  with respect to the reaction coordinate $\bs{\xi}(\bs{q})$ is
defined to be the \emph{effective potential} of $\bz = \bs{\xi}(\bs{q})$
\be
\label{eq:rcfree}
p(\bz) \propto \exp\left(-\beta A(\bz)\right)
\ee
Combining these two equations we see that:
\be
\label{eq:rcfree1}
 A(\bz) = -\beta^{-1}\log\int \exp\left(-\beta
V(\bs{q})\right)\delta(\bs{\xi}(\bs{q})-\bz)d\bs{q}
\ee

If $\hat A(\bz;\bt)$ is an estimate of  $A(\bz)$, we define a new probability distribution over
$\bs{q}$ by:
\be
\label{eq:rcqbias}
p(\bs{q}|\bt) \propto 
1_\mathcal{D}(\bs{\xi}(\bs{q}))
\exp\left(-\beta(V(\bs{q}) - \hat A(\xi(\bs{q});\bt))\right)
\ee
It is straight forward to see that under this new distribution for $\bs{q}$,
the pdf of $\bz$ becomes:
\be
\label{eq:rczbias}
p(\bz|\bt) = \int p(\bs{q}|\bt) \delta(\bs{\xi}(\bs{q})-\bz)d\bs{q}
\propto 
1_\mathcal{D}(\bz)
\exp\left(-\beta(A(\bz) - \hat A(\bz;\bt))\right)
\ee
This coincides with the expression in \refeq{eq:augmar} and therefore the ensuing derivations hold identically. From a practical point of view, sampling need only performed in the $\bq$ space and therefore the adaptive SMC schemes are employed to obtain particulate approximations of the density in \refeq{eq:rcqbias}. The only difference appears in the MCMC-based Rejuvenation step where the MALA sampler is employed only with regards to $\bq$. In particular the update of \refeq{eq:malaq} now becomes:
\be
\label{eq:malaq2}
\begin{array}{ll}
\bs{q}_{s}^{(i)}-\bs{q}_{s-1}^{(i)} & =\frac{\Delta t_q}{2} \nabla_{\bq} \pi_{\gamma_s}(\bs{q}_{s-1}^{(i)} ) +\sqrt{\Delta t_q} \bs{r}_q \\
& = -\frac{\beta \Delta t_q}{2} \left( \nabla_{\bq} V(\bs{q}_{s-1}^{(i)}) - \frac{\pa \hat{A}}{\pa \bz} \nabla_{\bq} \bs{\xi}(\bq) \right) +\sqrt{\Delta t_q} \bs{r}_q 
\end{array}
\ee

It is noted that, in contrast to some ABF methods which require second-order derivatives of $\bs{\xi}$ \cite{Henin2010}, the proposed technique only needs first-order derivatives. 
Finally, we point out that the ability of the proposed approach to provide efficiently estimates of parametrized  free energy surfaces (as in section \ref{sec:beta} with respect to the temperature $\beta$), can also be exploited in the reaction coordinate case by defining a joint density:
\be
\label{eq:densm}
p(\bq,\bz \mid \bt) \propto \exp \left\{-\beta \left( V(\bq) +\frac{\mu}{2} \parallel \bz-\bs{\xi}(\bq) \parallel^2 -\hat{A}_{\mu}(\bz;\bt)  \right) \right\}
\ee
where as in \cite{MARAGLIANO2006} an artificial spring with stiffness $\mu$ has been added. Clearly for $\mu \to \infty$ one recovers the aforementioned description, but for all other values of $\mu$ the formulation reduces to that of \refeq{eq:augmented} where in place of $V(\bq,\bs{z})$ we now have $V(\bq) +\frac{\mu}{2} \parallel \bz-\bs{\xi}(\bq) \parallel^2$. One can therefore obtain free energy surfaces for various $\mu$ values. For smaller $\mu$ the free energy would be flatter and in the extreme case of $\mu=0$ it would be constant. As $\mu$ increases, the complexities of the free energy surface would become pronounced. Hence by exploiting the idea of section \ref{sec:beta}, a sequence of problems parametrized by $\mu$ rather than $\beta$, can be constructed to   gradually move to larger $\mu$ values by using the free energy of the previous $\mu$  as an initial guess for the new one. The adaptive SMC scheme would ensure a smooth enough transition while retaining a good level accuracy for the approximations obtained.

\section{Numerical Examples}
\label{sec:examples}
\subsection{Two-Dimensional Toy Example}
\label{sec:ex_toy}

Consider a two-dimensional system \cite{voter1997method,lel07com}
with a single parameter $z$,
interacting with potential energy
$$
V(q;z) = \cos(2\pi z)(1 + d_1q) + d_2q^2
\label{eq:toy_pot}
$$
Assume that $q$ given $z$ and $\beta$ is distributed according to
$$
p(q|z,\beta) \propto \exp\left(-\beta V(q;z)\right)
$$
where $\beta$ is  also a fixed parameter that plays the role
of an inverse temperature.
We wish to calculate an approximation $\hat A(z)$ of the
free energy $A(z)$ on an interval $D=[-0.5,0.5]$.
The true free energy can be found analytically to be 
$$
A(z) = \cos(2\pi z) - \frac{d_1^2\cos(2\pi z)^2}{4d_2} + c
$$
where $c$ is a constant that depends upon the specific choice of the fixed
parameters. In what follows, we choose $c$ so that $A(-0.5) = 0$.

\begin{figure}[]
\subfigure[The potential energy $V(q_1,q_2)$ for $d_1=2,d_2=30$]{
\includegraphics[width=0.45\textwidth]{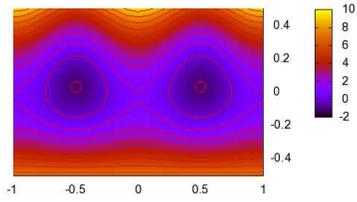}
\label{fig:toy_pot_con} 
} \hfill
\subfigure[The probability distribution $p(q_1,q_2|\beta)$ for $d_1=2,d_2=30,\beta=10$]{
\includegraphics[width=0.45\textwidth]{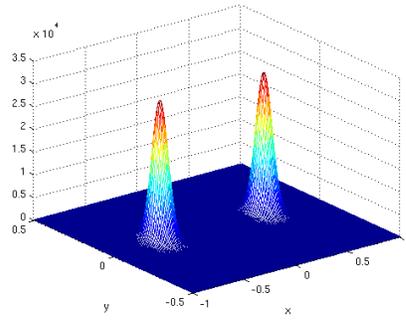}
\label{fig:toy_pdf}
}
\caption{Potential energy and pdf for the toy example of section \ref{sec:ex_toy}}
\label{fig:toy1}
\vspace{.5cm}
\end{figure}

\begin{figure}
\subfigure[K=1]{
 \label{fig:toy_evol100_k=1}
\psfrag{z}{$z $}
\psfrag{a}{$A(z)$}
\includegraphics[width=0.45\textwidth]{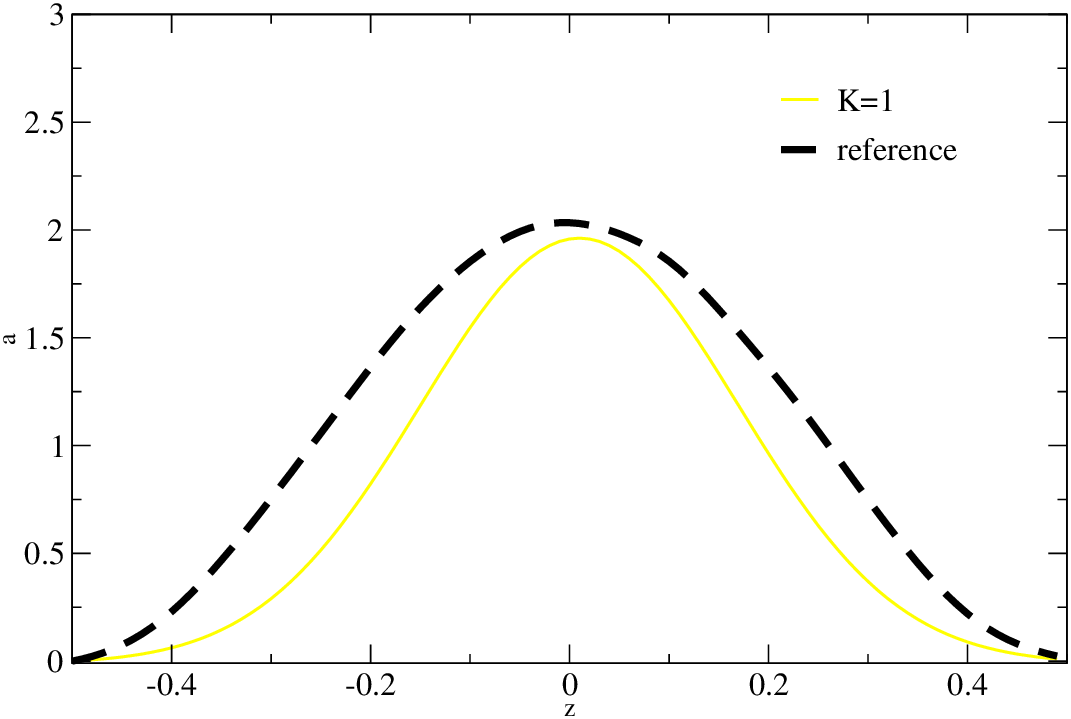}}
\hfill
\subfigure[K=2]{
 \label{fig:toy_evol100_k=2}
\psfrag{z}{$z $}
\psfrag{a}{$A(z)$}
\includegraphics[width=0.45\textwidth]{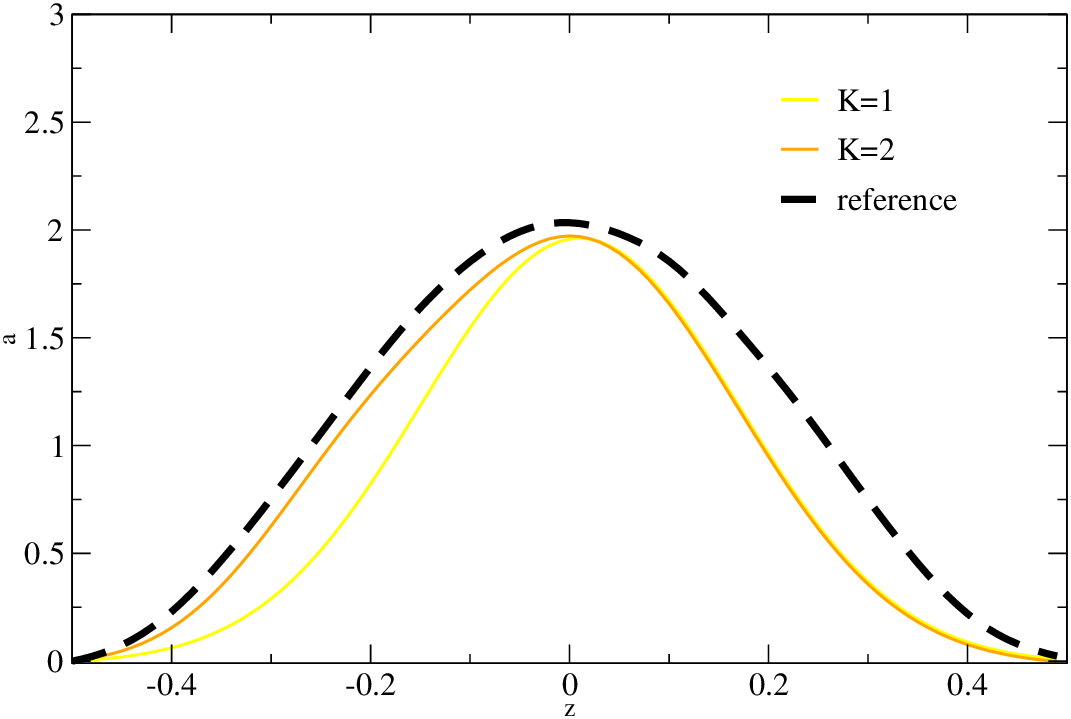}} 
\vfill
\vspace{.5cm}
\subfigure[K=3]{
\vspace{3cm}
 \label{fig:toy_evol100_k=3}
\psfrag{z}{$z $}
\psfrag{a}{$A(z)$}
\includegraphics[width=0.45\textwidth]{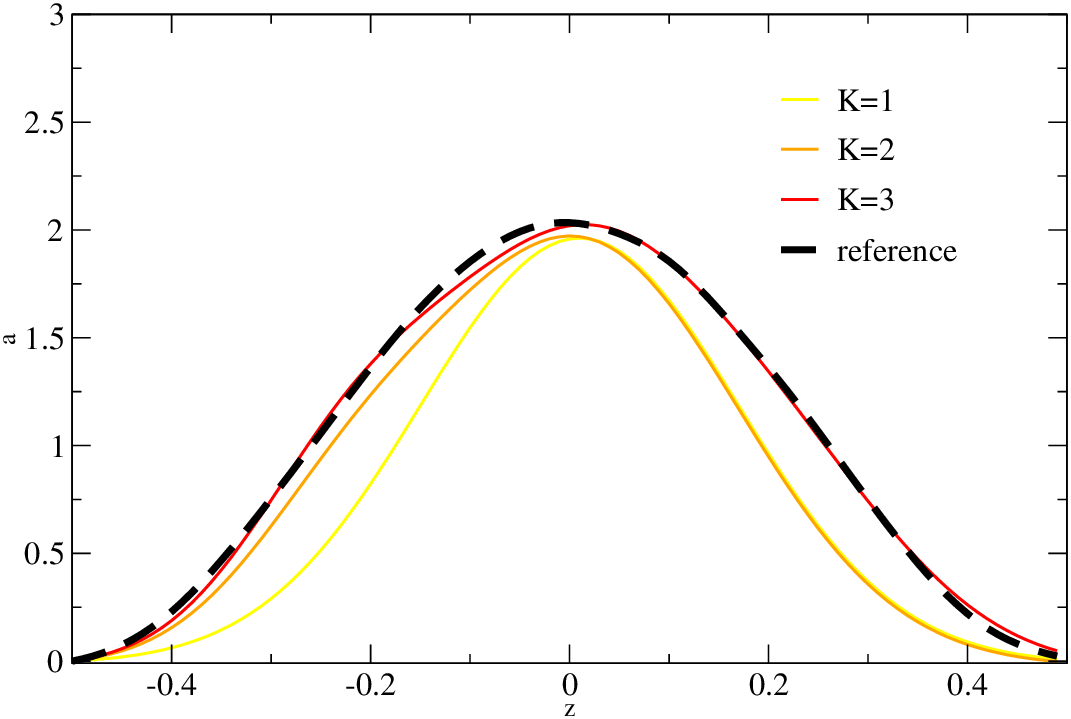}}
\hfill
\subfigure[K=8]{
 \label{fig:toy_evol100_k=8}
\psfrag{z}{$z $}
\psfrag{a}{$A(z)$}
\includegraphics[width=0.45\textwidth]{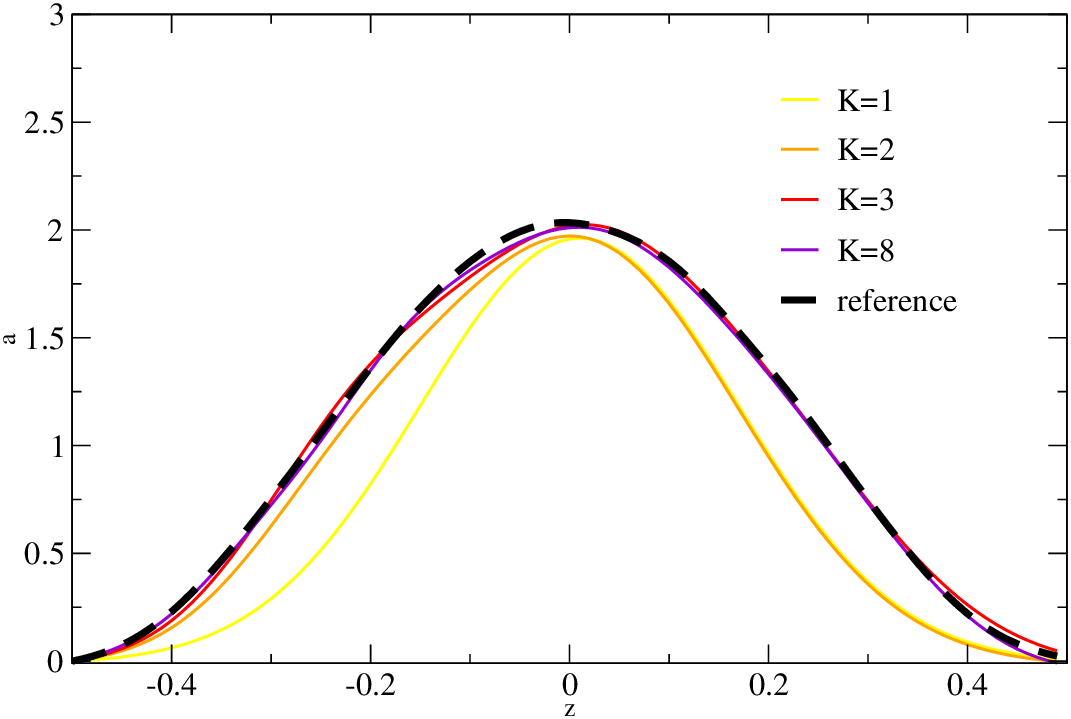}}
\caption{Evolution of free energy estimates at various kernel numbers $K$ when using $n=100$ particles in the adaptive SMC scheme}
 \label{fig:toy_evol100}
\vspace{.5cm}
\end{figure}

\begin{figure}
\subfigure[K=1]{
 \label{fig:toy_evol10000_k=1}
\psfrag{z}{$z $}
\psfrag{a}{$A(z)$}
\includegraphics[width=0.45\textwidth]{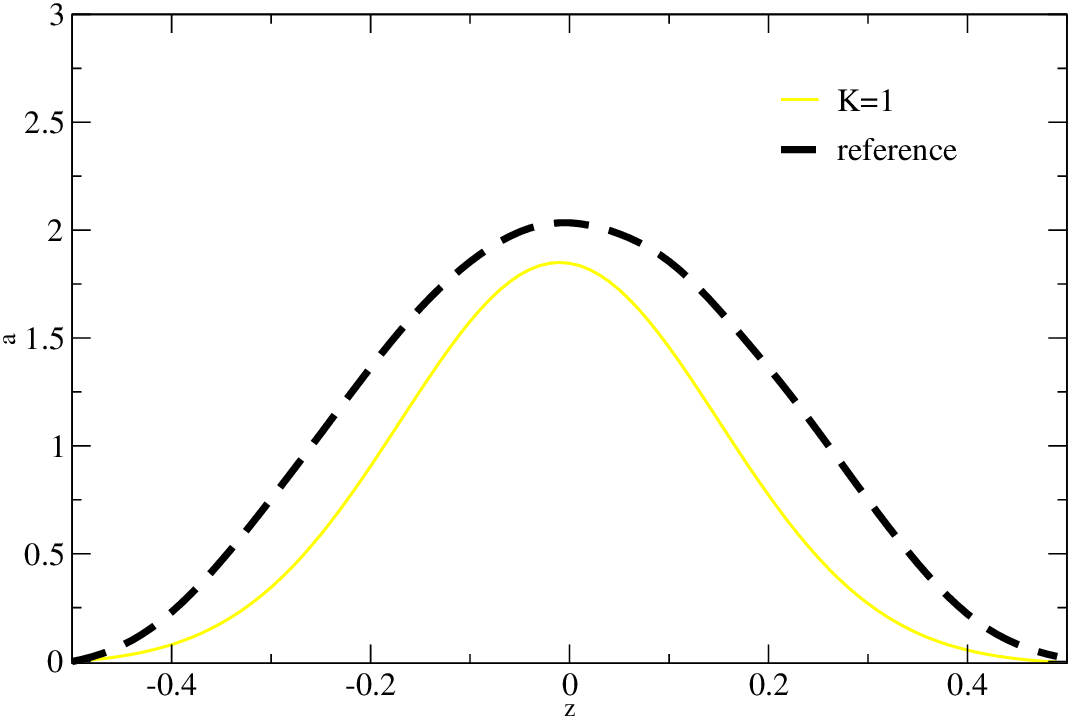}}
\hfill
\subfigure[K=2]{
 \label{fig:toy_evol10000_k=2}
\psfrag{z}{$z $}
\psfrag{a}{$A(z)$}
\includegraphics[width=0.45\textwidth]{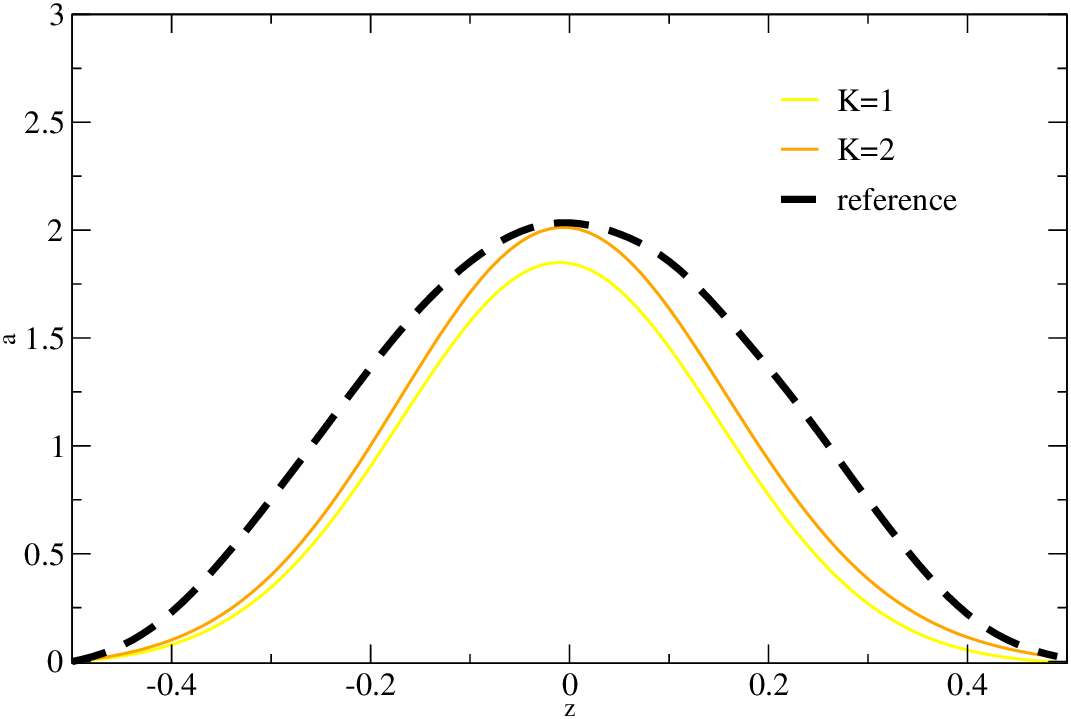}} \\
\vfill
\vspace{.5cm}

\subfigure[K=3]{
 \label{fig:toy_evol10000_k=3}
\psfrag{z}{$z $}
\psfrag{a}{$A(z)$}
\includegraphics[width=0.45\textwidth]{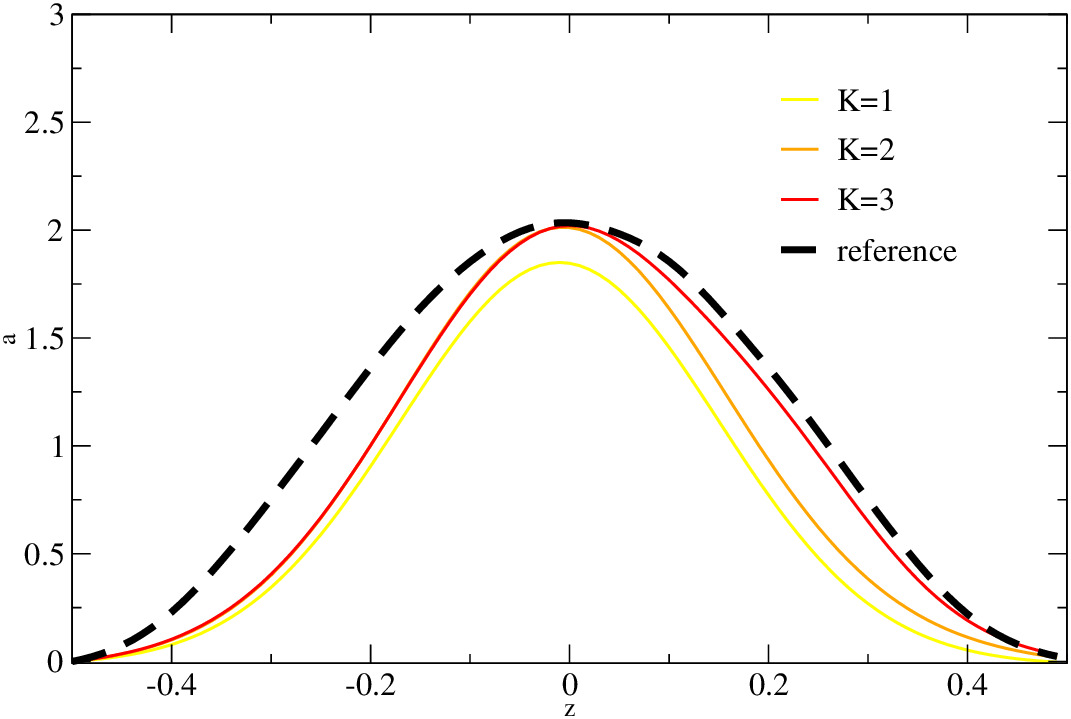}}
\hfill
\subfigure[K=8]{
 \label{fig:toy_evol10000_k=8}
\psfrag{z}{$z $}
\psfrag{a}{$A(z)$}
\includegraphics[width=0.45\textwidth]{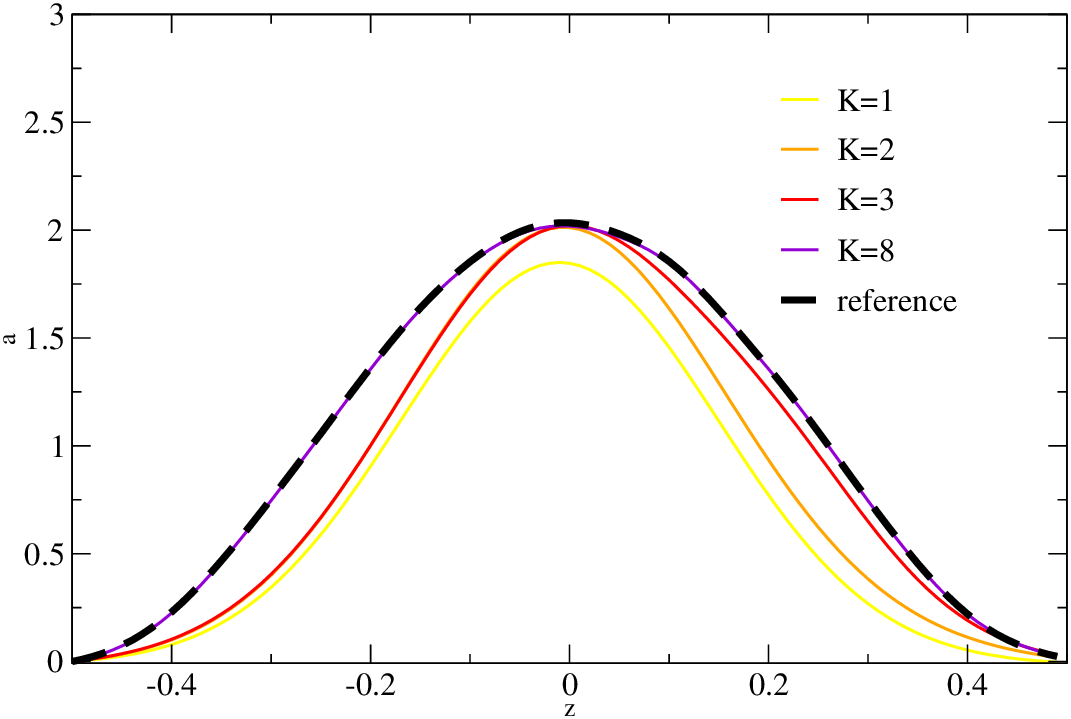}}
\caption{Evolution of free energy estimates at various kernel numbers $K$ when using $n=10,000$ particles in the adaptive SMC scheme}
 \label{fig:toy_evol10000}
\vspace{.5cm}
\end{figure}

To demonstrate our method in this simple example we used $d_1 = 2, d_2 = 30$. The potential energy $V(q;z)$
for this choice of the parameters is depicted in Figure \ref{fig:toy_pot_con}. We fix
the inverse temperature to $\beta=10$. As shown in Figure \ref{fig:toy_pdf}, the distribution
is bimodal with a big region of practically zero probability separating the two modes.
Hence, metastability along the parameter $z$ is apparent.
The performance of the proposed method with respect to the number of particles used in the adaptive SMC scheme  is depicted in Figures \ref{fig:toy_evol100} and  \ref{fig:toy_evol10000} which show the evolution of the estimated free energy landscape with $n=100$  and $n=10,000$ particles respectively. In both cases the method is capable of capturing the correct  characteristics of the reference solution  and as expected the noise in the computations is less when the number of particles is larger. 
In both cases  the Robbins-Monro learning series is picked to be $\eta_m =  m^{-p}$ 
with $p=0.6$ and the learning rate $\lambda = 0.1$. 

\begin{figure}
\vspace{.5cm}
\subfigure[First three kernels $K_j(.)$  picked by the greedy optimization scheme to illustrate selection of location and bandwidth parameters ($n=10,000$)]{
 \label{fig:kernels}
\psfrag{z}{$z$}
\psfrag{a}{}
\includegraphics[width=0.45\textwidth]{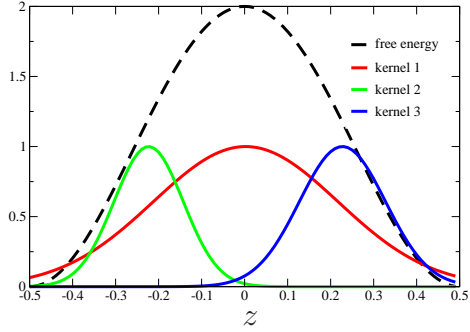}}\hfill
\subfigure[Log absolute weights ($\log \mid \theta_j \mid $) of the first $10$ kernels added. The $\theta$ value of the first kernel is over one order of magnitude larger than the rest]{
 \label{fig:logweights}
\psfrag{theta}{$\log \mid \theta_j \mid $}
\psfrag{kernel}{\small kernel number $j$}
\includegraphics[width=0.45\textwidth]{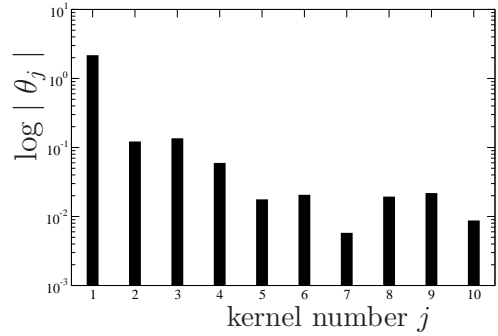}}
\caption{Kernels selected and kernel weights obtained with $n=10,000$ particles}
\vspace{.5cm}
\end{figure}

\begin{figure}[!ht]
\vspace{.5cm}
\psfrag{kl}{KL divergence}
\centering
\psfrag{iter}{iterations}
\psfrag{k1}{\scriptsize $K=1$}
\psfrag{k2}{\scriptsize $K=2$}
\psfrag{k3}{\scriptsize $K=3$}
\psfrag{k4}{\scriptsize $K=4$}
\psfrag{k5}{\scriptsize $K=5$}
\psfrag{k6}{\scriptsize $K=6$}
\psfrag{k7}{\scriptsize $K=7$}
\includegraphics[width=0.75\textwidth]{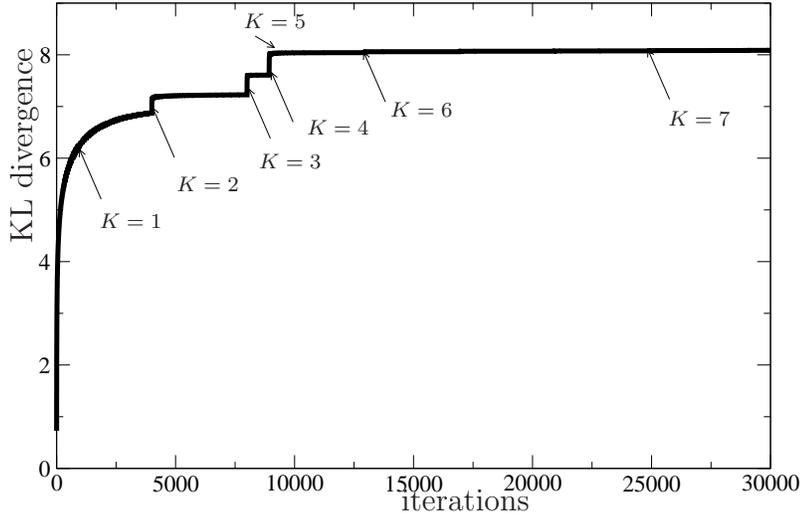}
\caption{Evolution of  the reduction in the KL divergence at various $K$  with respect tot the total number of iterations performed. Upon addition of each kernel a reduction $\Delta_K$ (\refeq{eq:kl_impr}) is achieved which becomes progressively smaller. }
\label{fig:toy_kl}
\end{figure}

Figure \ref{fig:kernels} shows the first three kernels selected by the greedy scheme of section \ref{sec:kernel}. Figure  \ref{fig:logweights} depicts the log-values of the kernel weights $\{\theta_j\}_{j=1}^K$ 
 which clearly demonstrate the ability of the proposed approach to provide sparse approximations. The first kernel selected  has the greatest weight and hence it contains
the majority of the information about the free energy curve. The rest of the kernels are progressive
corrections of the estimate given by the first kernel. This conclusion is also supported by the result of  Figure \ref{fig:toy_kl} which shows the evolution of the reduction in the KL divergence with respect to the total number of iterations as quantified by adding the $\Delta_{K+1}$ in \refeq{eq:kl_impr}. Clearly the first kernel offers the greatest KL gain ($\Delta_1$) and further kernel additions offer progressively smaller reductions in the KL divergence.

\subsection{WCA Dimer}
\label{sec:ex_dimer}
We consider $N=16$ atoms in a two-dimensional fully periodic box of side $l$ which interact with a purely repulsive WCA pair potential \cite{lel07com}:
$$
V_{\text{WCA}}(r) = 
\begin{cases}
4\epsilon\left[\left(\frac{\sigma}{r}\right)^{12} 
					- \left(\frac{\sigma}{r}\right)^6\right]
					+ \epsilon&, \text{if\;\;} r\ge r_0\\
0&, \text{otherwise}
\end{cases}
$$
where $\sigma$ and $\epsilon$ give the length and energy scales respectively.
Two of these atoms (say atoms 1 and 2) are assumed to form a dimer and their interaction is described instead with a double well potential:
$$
V_S(r) = h\left[1 - \frac{(r-r_0-w)^2}{w^2}\right]
$$
where $h,w,r_0$ are fixed parameters and $r$ the distance between them.
This potential has two equilibrium points $r_0$ and $r_0 + 2w$. The barrier
separating the two equilibria is $h$.

Let $\bs{q}=(\bs{q}_1,\bs{q}_2,\dots,\bs{q}_N)$ with $\bs{q}_i\in\RR^2$ denoting 
the position of atom $i$. The potential energy of the system is:
$$
V(\bs{q}) = V_S(|\bs{q}_1-\bs{q}_2|) +
 \sum_{i=1}^2\sum_{j=3}^N V_{\text{WCA}}(|\bs{q}_i-\bs{q}_j|) + 
 \sum_{2<i<j}V_{\text{WCA}}(|\bs{q}_i-\bs{q}_j|)
$$
We consider an NVT ensemble (the volume $V$ is determined by the side of the box $l$).
The probability distribution of the atomic positions $\bs{q}$ is
$$
p(\bs{q}|\beta) \propto \exp\left(-\beta V(\bs{q})\right)
$$
where $\beta = \frac{1}{k_B T}$, $k_B$ is the Boltzman constant  and $T$ the temperature
of the system. Under these assumptions atoms 1 and 2 will form a dimer with two
equilibrium lengths. An \emph{effective potential} of the dimer length in the 
presence of the other atoms is given by the free energy $A(r)$ with respect to the 
reaction coordinate
$$
z = \xi(\bs{q}) = \parallel\bs{q}_1-\bs{q}_2\parallel_2
$$
where $||\cdot||_2$ is the Euclidean norm of $\RR^2$.

\begin{figure}[]
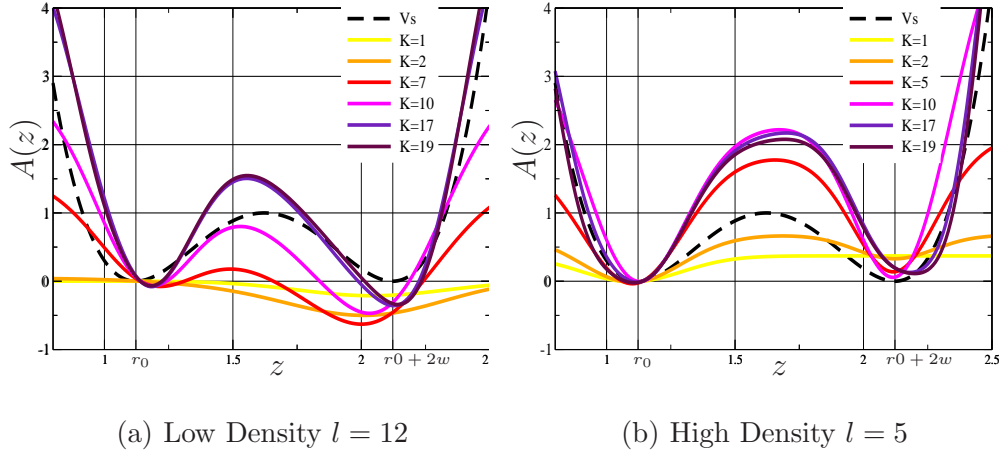

\psfrag{z}{$z$}
\psfrag{a}{$A(z)$}
\psfrag{r0}{\tiny $r_0$}
\psfrag{r0w}{\tiny $r0+2w$}
	\begin{center}
		\subfigure[Low Density $l=12$]{
\includegraphics[width=0.45\textwidth,height=5cm]{FIGURES/evolution_dimer_l=12.eps}		
	\label{fig:dimer_low}
		}
		\subfigure[High Density $l=5$]{
\includegraphics[width=0.45\textwidth,height=5cm]{FIGURES/evolution_dimer_l=5.eps}		
			\label{fig:dimer_high}
		}
	\end{center}
	\caption{The free energy of the dimer at two different densities compared with $V_S(r)$.
	Notice that at low density \subref{fig:dimer_low} the right well becomes the most probable. This situation
	is reversed at high density \subref{fig:dimer_high}.
	}
	\label{fig:dimer}
\vspace{.5cm}
\end{figure}

We calculate $A(z)$ using our scheme for two different box sizes (densities):
$l=4$ (high density) and $l=12$ (low density). The parameters are set to
$N=16$ atoms, $\beta=1, \epsilon=1, \sigma = 1, h = 1, w = 0.5$. 
We employed $n=500$ particles and the Robbins-Monroe learning series is again $\eta_m = m^{-p}$ with
 $p=0.501$ and $\lambda=0.1$. The resulting free energy curves at various stages of the estimation process with increasing number of kernels  are shown
in Figure \ref{fig:dimer}. We notice that at {\em low density} i.e. when the box size is  $l=12$ (Figure \ref{fig:dimer_low}), 
the equilibria move to the right with the well closest to $r_0+2w$ becomes the most probable.
Furthermore the free energy barrier is slightly decreased as compared to the {\em high density} case when  $l=5$ (Figure \ref{fig:dimer_high}). Under these conditions the  equilibria
move to the left and the well closest to $r_0$  becomes the most probable.

\newpage

\subsection{$38$-Atom Lennard-Jones Cluster ($LJ_{38}$)}
\label{sec:ex_cluster}
\begin{figure}
 \subfigure[$Q_4 \approx 0.01$]{
\label{fig:icosa}
\includegraphics[width=0.45\textwidth]{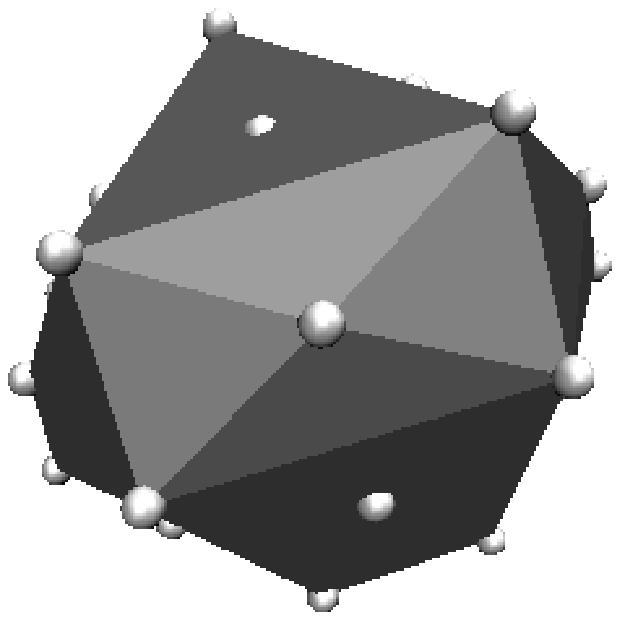}}
\subfigure[$Q_4 \approx 0.19$ (truncated octahedron)]{
\label{fig:octa}
\includegraphics[width=0.45\textwidth]{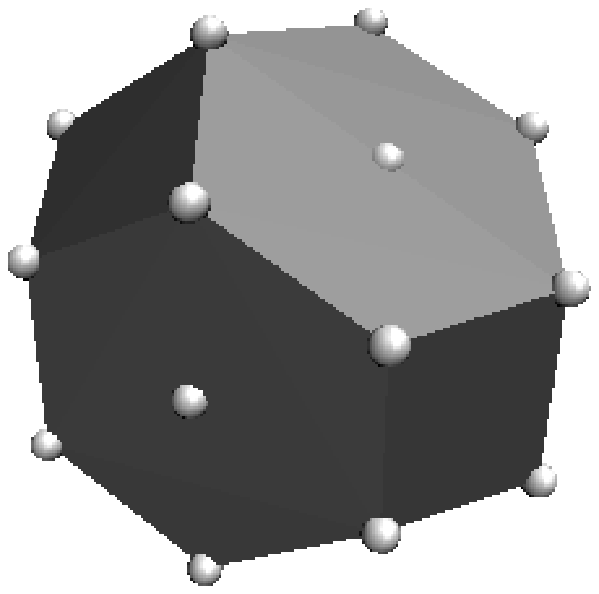}}
\caption{Indicative metastable states corresponding to the two wells of the free energy landscape with respect to order parameter $Q_4$ (\refeq{eq:order})}
\vspace{0.5cm}
\end{figure}

We consider a $38$-atom cluster in $3$-dimensional space with pairwise interactions
given by the Lennard-Jones potential:
\be
V_{\text{LJ}}(r) = \left[
\left(\frac{\sigma}{r}\right)^{12} - \left(\frac{\sigma}{r}\right)^6
\right]
\ee
with $\epsilon$ and $\sigma$ playing the role of energy and length scale respectively.
Let the Cartesian coordinates of the system be
\be
\bq = \left(\bq_1,\dots,\bq_{38}\right), \bq_i\in\RR^3
\ee
Then the potential energy of the system is
$$
V(\bq) = \frac{1}{2}\sum_{i<j} V_{\text{LJ}}\left(|\bq_i-\bq_j|\right)
$$
Finally we assume that the particles follow an NVT distribution of the form 
$$
p(\bq|\beta) \propto \exp\left\{-\beta V(\bq) \right\}
$$
where $\beta = 1/k_BT$.
At zero temperature the system is known to have a global minimum yielding an
FCC truncated octahedron (Figure \ref{fig:octa}). The second and third lower energies
give incomplete Mackey icosahedra. Furthermore there is a big number of 
liquid-like local minima (\cite{Doye1999,Calvo2000}).

\begin{figure}
\psfrag{Q4}{$Q_4$}
\psfrag{E}{$E$}
\subfigure[$T=0.21$]{
 \label{fig:lj38_21}
\includegraphics[width=0.45\textwidth]{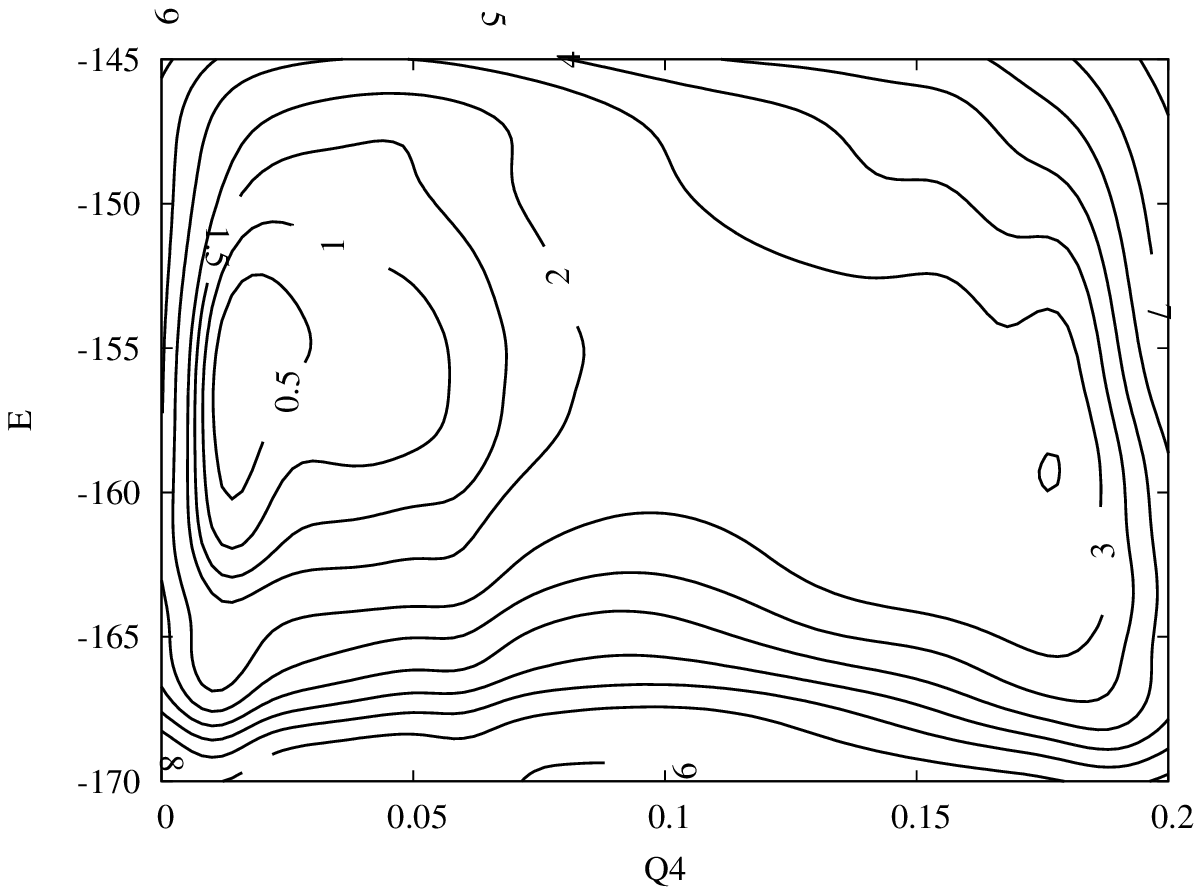}}
\hfill
\subfigure[$T=0.17$]{
 \label{fig:lj38_17}
\includegraphics[width=0.45\textwidth]{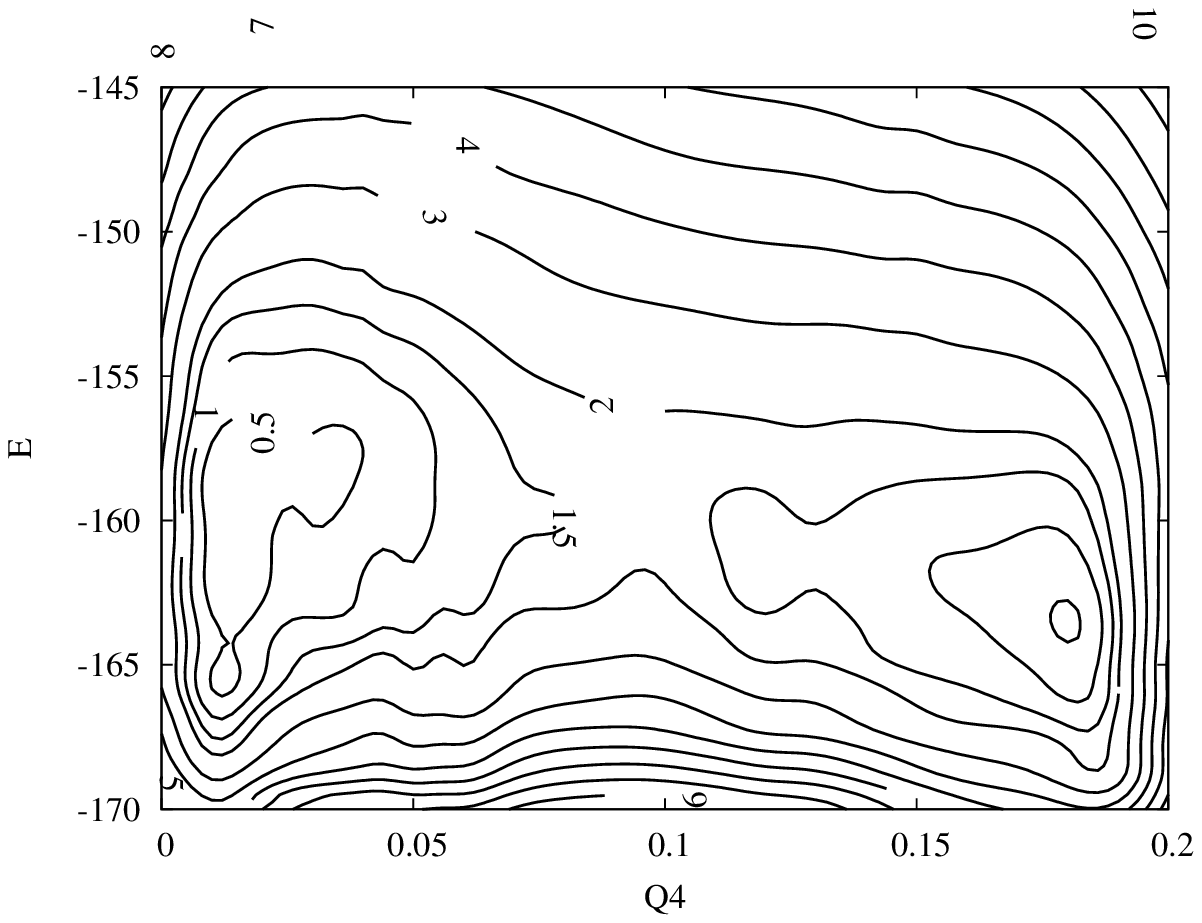}} \\
\subfigure[$T=0.14$]{
 \label{fig:lj38_14}
\includegraphics[width=0.45\textwidth]{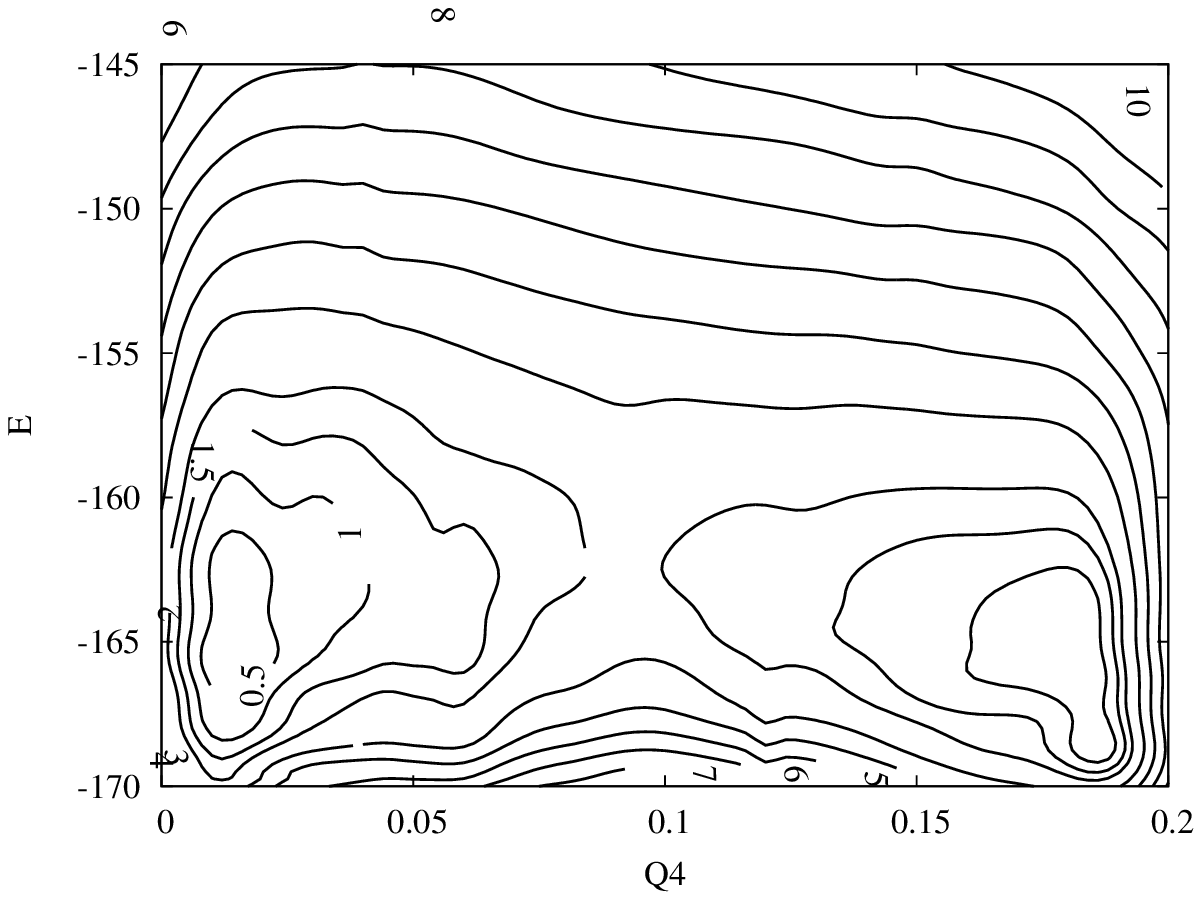}}
\hfill
\subfigure[$T=0.11$]{
 \label{fig:lj38_11}
\includegraphics[width=0.45\textwidth]{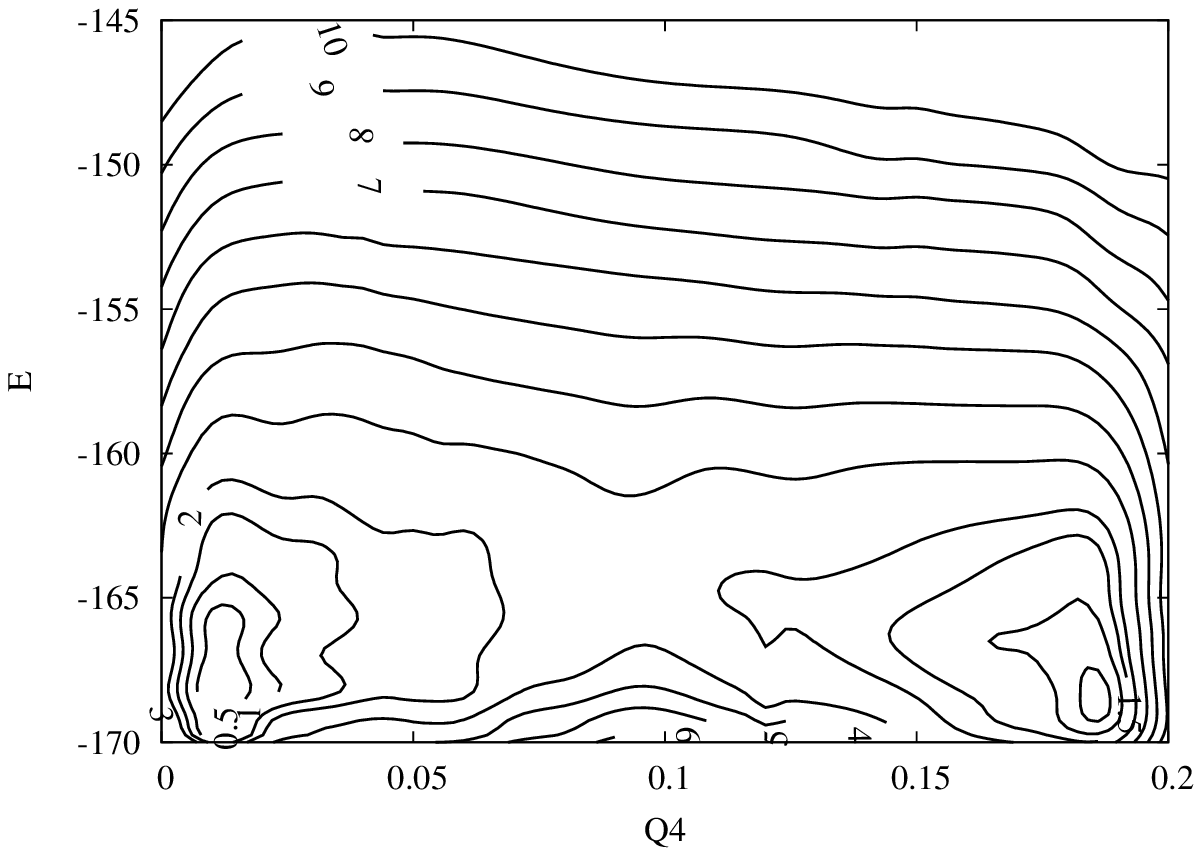}}
\caption{Free energy contours $A(Q_4,E)$ with respect to the two reaction coordinates $Q_4$ (x-axis) and $E$ (y-axis) at various temperatures for $LJ_{38}$ }
 \label{fig:lj38}
\end{figure}

Consider the family of order parameters initially introduced in \cite{Steinhardt1983}:
\be
\label{eq:order}
Q_l = \left(\frac{4\pi}{2l+1}\sum_{m=-l}^l|\bar{Q}_{lm}|^2\right)^2
\ee
with
$$
\bar{Q}_{lm} = \frac{1}{N_b}\sum_{r_{ij} < r_0} Y_{lm}(\theta_{ij},\phi_{ij})
$$
where the sum is over all the $N_b$ pairs of atoms with $r_{ij} = |\bq_i-\bq_j| < r_0$,
$Y_{lm}(\theta,\phi)$ is a spherical harmonic, while $\theta_{ij}$ and $\phi_{ij}$ 
are the polar and azimuthal angles of a bond vector with respect to an arbitrary 
coordinate system. In \cite{Calvo2000} it is shown that for $l=4$, $Q_4$ can distinguish 
the FCC structure but not the icosahedral and liquid-like minima (Figure \ref{fig:icosa}). However, if one also
considers the energy the two structures are well-separated. Hence, we define the two 
dimensional reaction coordinate:
$$
\xi(\bq) = \left(Q_4(\bq), V(\bq)\right)
$$
we compute the free energy 
$$
A(Q_4,E) = \beta^{-1}\int \exp\left\{-\beta V(\bq)\right\}\delta(Q_4-Q_4(\bq))\delta(E-V(\bq))d\bq
$$
over the domain
$$
D = [0,0.2]\times[-175\epsilon, -145\epsilon]
$$
for a temperature range $k_BT = 0.21$ to $k_BT = 0.091$ using the tempering scheme described
in Section \ref{sec:beta}. We employ $n=100$ particles and $10$ MCMC/Rejuvenation steps per particle. At each  $\beta = k_B T$, the Robbins-Monro learning series
was adjusted to $\eta_m =  m^{-p}$ with $p = 0.501$ and a learning rate $\lambda= 0.1/\beta$. The adaptive SMC scheme automatically determined $260$ intermediate steps/distributions in order to cover the whole range of the aforementioned temperatures. 
While the time step $\Delta t_q$ employed in the MALA sampler was adaptively  adjusted as discussed previously  and took values between $10^{-4}$ (low temperatures) and $7\times 10^{-4}$ (high temperatures).
The very first step, at $T=0.21$ ($\beta=4.76$) required $12,000$ optimization iterations to converge with a cost of approximately $7.2\times 10^{5}$ time steps per particle. It is emphasized that due to the embarrassingly parallelizable nature of the SMC scheme employed , each particle can be simulated in a different CPU, largely independently of the rest.
The sequence of intermediate $\beta$'s determined automatically by the scheme discussed in section \ref{sec:beta} is depicted in Figure \ref{fig:lj38_beta}.
The similarity of the free energy surfaces at neighboring temperatures allowed us to converge with, on average, $800$ optimization iterations at each intermediate $\beta$. The overall cost  was $2.4 \times 10^4$ time steps per particle, i.e. equivalent to $1/30$ of the cost for calculating the free energy from scratch at the initial $\beta$. 
The free energy surfaces computed are depicted in Figure \ref{fig:lj38} at four indicative temperatures. The number of kernels selected by the algorithm  varied between $90$ and $120$.
As it has been reported in previous studies \cite{Calvo2000}, we identified two metastable states at $Q_4\approx 0.01$ which corresponds to the truncated octahedron and at $Q_4 \approx 0.19$ which corresponds to the icosahedron. The latter becomes more pronounced at lower temperatures.

\begin{figure}
\vspace{.5cm}
\centering
\psfrag{iteration}{intermediate step number}
\psfrag{beta}{$\beta$}
\psfrag{T}{$T$}
\includegraphics[width=0.75\textwidth]{FIGURES/beta_sequence.eps}
\caption{Sequence of intermediate $\beta$'s identified by the scheme discussed in section \ref{sec:beta} for the $LJ_{38}$ cluster. The free energy landscape was calculated at each of these temperatures by efficiently updating the free energy surface at the previous step.}
\label{fig:lj38_beta}
\end{figure}

\begin{figure}
\vspace{.5cm}
\centering
\psfrag{Q4}{$Q_4$}
\psfrag{az}{$A(Q_4)$}
\includegraphics[width=0.75\textwidth]{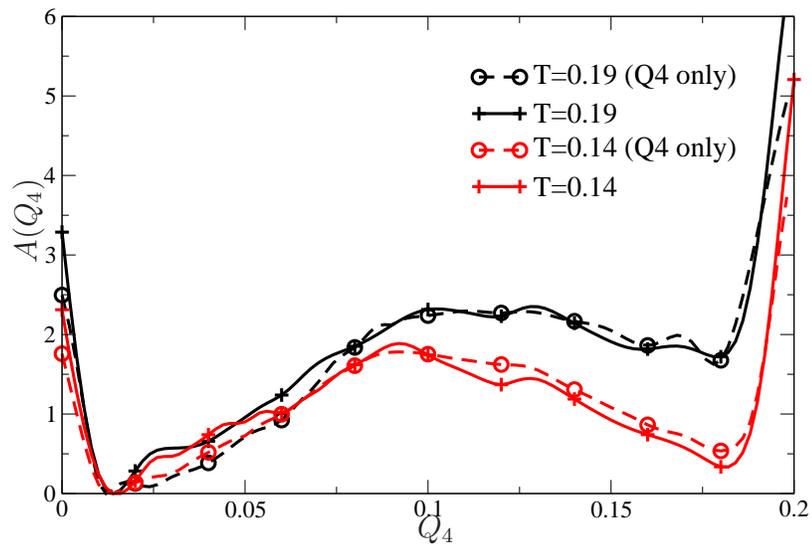}
\caption{Free energy profiles $A(Q_4)$  obtained using the proposed scheme while using only reaction coordinate (dashed lines) and by integrating the two dimensional free energy surface i.e. $A(Q_4)= -\beta^{-1} \log \int e^{-\beta A(Q_4,E)} dE$ (solid lines) for two temperatures $T=0.19$ and $T=0.14$}
\label{fig:lj38_com}
\vspace{.5cm}
\end{figure}

In order to assess the quality of the results in two dimensions we also calculated the free energy profile using only $Q_4$ as the reaction coordinate (Figure \ref{fig:lj38_com}). We also performed a numerical integration of the two-dimensional free energy surface i.e. by computing $A(Q_4)= -\beta^{-1} \log\int e^{-\beta A(Q_4,E)} dE$ with respect to the second reaction coordinate. The two free-energy curves are depicted in Figure \ref{fig:lj38_com} where good agreement is observed at two different temperatures.

\section{Conclusions}
In summary, the proposed method provides a unifying framework for estimating the free energy function simultaneously with biasing the dynamics. The minimization of the Kullback-Leibler divergence in the extended space provides rigorous convergence bounds and diagnostics. It requires minimal  adjustment of parameter  values a priori (basically only the learning rate $\lambda$ and convergence tolerances) as it is adaptive and automatically promotes sparse representations of the free energy surface.  
It offers several possibilities for further improvements by considering different optimization schemes (e.g. noisy conjugate gradients) and employing different basis functions (e.g. wavelets).
Its sequential nature allows the efficient computation  of a family of free energy surfaces at different temperatures.
We believe that these features make the proposed approach suitable  to calculate the free energy of systems
more physically challenging  than the ones discussed in this paper.

\textit{Acknowledgement} We are much indebted to Dr. Florent Calvo for providing us with
the Q4-gradient subroutine. This   work was supported by the OSD/AFOSR MURI'09 award to Cornell University on uncertainty quantification
 \\

\clearpage
\newpage

\bibliographystyle{plain}
\bibliography{paper}

\end{document}